\documentclass[apj, twocolumn]{aastex62}
\shorttitle{Delay Time Distributions of Type Ia Supernovae}
\shortauthors{Strolger et al.}

\usepackage{natbib}
\usepackage{nicefrac}
\usepackage{ulem}
\usepackage{xcolor}
\usepackage{topcapt}
\usepackage{amssymb,amsmath}
\begin{document}
\title{Delay Time Distributions of Type Ia Supernovae From Galaxy and Cosmic Star Formation Histories}
\author[0000-0002-7756-4440]{Louis-Gregory Strolger}
\affiliation{Space Telescope Science Institute, 3700 San Martin Drive, Baltimore MD 21218, USA}
\correspondingauthor{Louis-Gregory Strolger}
\email{strolger@stsci.edu}

\author{Steven A.~Rodney}
\affiliation{Department of Physics and Astronomy, University of South Carolina, 712 Main St., Columbia, SC 29208, USA}

\author{Camilla Pacifici}
\affiliation{Space Telescope Science Institute, 3700 San Martin Drive, Baltimore MD 21218, USA}

\author{Gautham Narayan}
\affiliation{University of Illinois Urbana-Champaign. 1002 W. Green Street, Urbana, IL 61801, USA}

\author{Or Graur}
\affiliation{Center for Astrophysics | Harvard \& Smithsonian, 60 Garden Street, Cambridge, MA 02138, USA}
\affiliation{Department of Astrophysics, American Museum of Natural History, Central Park West and 79th Street, New York, NY 10024, USA}
\affiliation{NSF Astronomy and Astrophysics Postdoctoral Fellow.}

\begin{abstract}
We present analytical reconstructions of type Ia supernova (SN~Ia) delay time distributions (DTDs) by way of two independent methods: by a Markov chain Monte Carlo best-fit technique comparing the volumetric SN~Ia rate history to today's compendium cosmic star-formation history, and secondly through a maximum likelihood analysis of the star formation rate histories of individual galaxies in the GOODS/CANDELS field, in comparison to their resultant SN~Ia yields. {We adopt a flexible skew-normal DTD model, which could match a wide range of physically motivated DTD forms.} We find a family of solutions that are essentially exponential DTDs, similar in shape to the $\beta\approx-1$ power-law DTDs, but with more delayed events ($>1$ Gyr in age) than prompt events ($<1$ Gyr). {Comparing these solutions to delay time measures separately derived from field galaxies and galaxy clusters, we find the skew-normal solutions can accommodate both without requiring a different DTD form in different environments}. These model {fits} are generally inconsistent with results from single-degenerate binary population synthesis models, and are seemingly supportive of double-degenerate progenitors for most SN~Ia events. 

\end{abstract}

\section{Introduction}
The understanding of cosmic type Ia supernova (SN~Ia) rates has critical importance to understanding galaxy evolutionary feedback mechanisms, cosmic iron enrichment and $\alpha$-process element enrichment histories \citep[see][]{Maoz:2017ck}, and perhaps most importantly, constraining the physical mechanisms of SN~Ia progenitors, and therefore providing some constraint on the systematic uncertainties of dark energy. However, determining precise SN~Ia rates and rate histories, and establishing the connections of those rates to host (and cosmic) properties has been a long slog. 

In tracing the cosmic (or volumetric) rate history, the first precise measures of the local ($z\sim0$) rate came in the early 1990s \cite[see ][]{Cappellaro:1993qm,Cappellaro:1999}, and the first measures beyond the local Hubble flow came in the early 2000s, many as collateral results of dark energy experiments~\citep{Riess:1998,Perlmutter:1999}. This trend of collateral benefit has continued since, leading to a vast collection of volumetric rates over various redshift ranges, and from various groups, as shown in Figure~\ref{fig:sn1a_rates} (the individual measures are shown in Appendix~\ref{sec:ratemeas}). 

\begin{figure*}[t]
   \centering
   \includegraphics[width=6.1in]{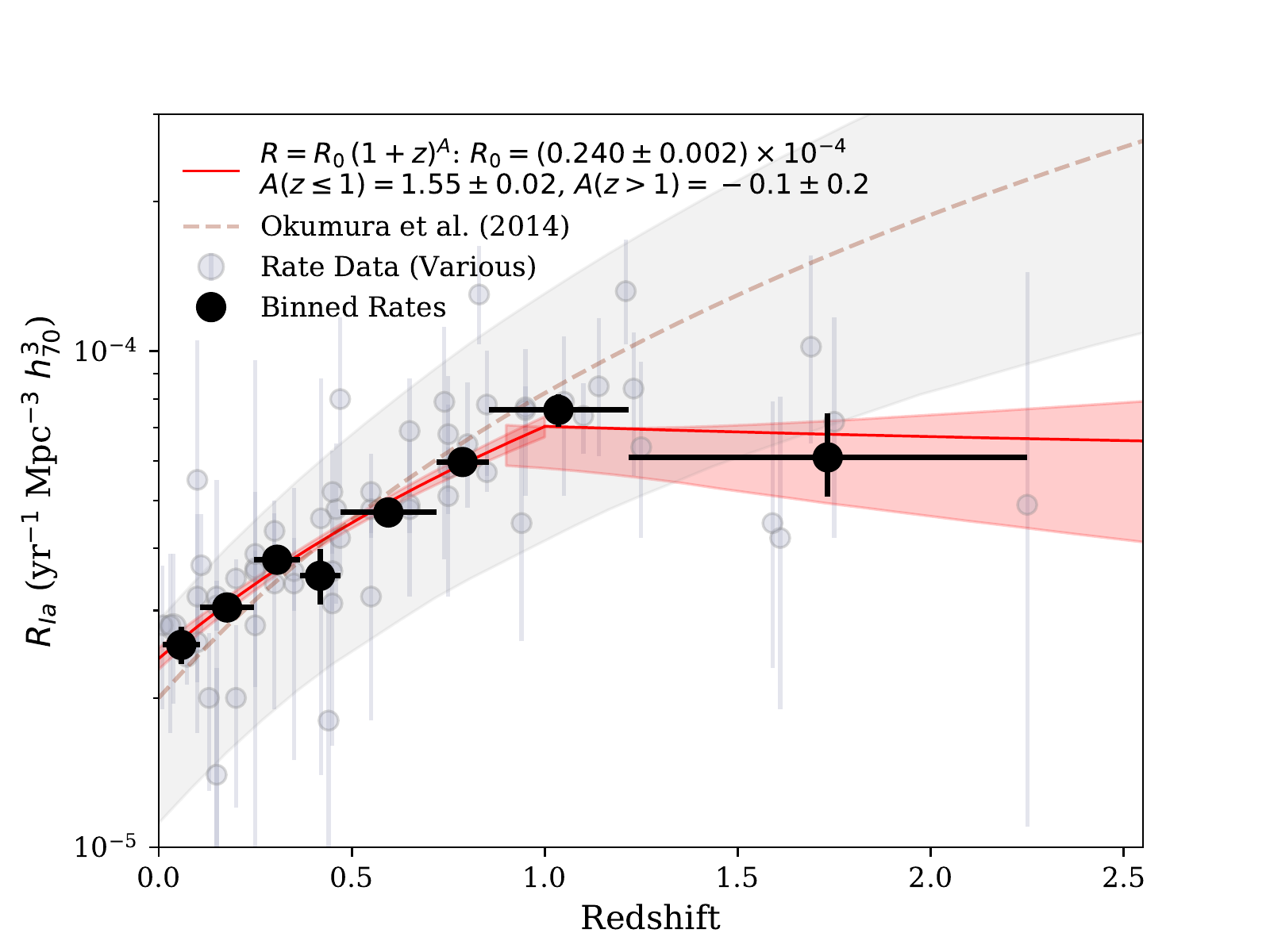}
   \caption{\footnotesize Type Ia supernova volumetric rate measures from various sources in the literature (gray points, see Table~\ref{tab:sn1a_rates} for their sources), and binned (black points, see Table~\ref{tab:sn1a_bin}), largely for illustration. The solid red lines show a broken power-law fit to the data in redshift space. The dashed red line (and associated uncertainty region, in gray) is from~\cite{Okumura:2014}.}
   \label{fig:sn1a_rates}
\end{figure*}

Not all rate measures have been in agreement with one another (see the large scatter just below $z\sim0.5$), with the reasons as to why ranging from statistical variation, differences in the treatment of declining SNe, and ultimately differences in the assessment of effective survey duration, via modeling and simulation. It is left to a future study to attempt a reanalysis of some (or all) of the reported rate measures in at least a self-consistent assessment to reduce some of the unreported systematic uncertainties. For the time being, it is probably best to consider each published rate as a valid measure that may (or may not) have misestimated uncertainties. 

From this rate history and a comparison to the cosmic star formation history, one can reconstruct (or infer) the distribution of times from SN progenitor formation to explosion, assuming the mechanism is ubiquitous enough that it can be characterized by a singular distribution of delay times. It has been long expected that this `delay time distribution' (or DTD) will distinguish between single-degenerate \citep{Whelan:1973,Nomoto:1982vh} and double-degenerate \citep{Iben:1984, Webbink:1984} models, depending on what the models for each of these scenarios would expect (see Section~\ref{sec:dtd} for more details). {In the absence of definitive evidence favoring one of these traditional models, many variations and alternative models have also been evaluated \citep{Maoz:2014fj, Livio:2018rz}.}  \cite{Totani:2008} was among the first to show that the DTD is well described as a power-law function with time, with a slope of approximately -1, and generically consistent with double-degenerate scenarios.

Another method in reconstructing delay times is to use an analysis of the star-formation histories in individual host galaxies~\citep{Brandt:2010, Maoz:2011, Maoz:2012a}, under the same assumptions as above. Rather than using bulk star formation properties, averaged in large temporal bins (or by redshift), this method uses spectral analysis tools to reconstruct individual galaxy star formation histories. One can then determine the contributions (over broad periods or time bins) from a DTD that maximizes the likelihood that the hosts would produce SNe Ia in the duration of a survey. The results of these studies to date are (A) dependent on the presumed power-law shape of the DTD model, (B) tend to have different power-law slopes in galaxy cluster and field environments {(although not always, see Section~\ref{sec:discussion})}, independent of redshift, and (C) have not been tested on high-$z$ ($z\gtrsim1$) field galaxies. 

In this paper, we present an analysis of DTDs using these two separate methods, from a comparison of volumetric SN~Ia rates to cosmic star formation rates (Section~\ref{sec:rates}), and a maximized likelihood method from SN~Ia host star formation histories (Section~\ref{sec:sfh}). In Section~\ref{sec:discussion} we discuss the results from each of these methods and put them into context with results from other work, and into context with binary synthesis models. Throughout this manuscript we assume $H_0=70$ km s$^{-1}$ Mpc$^{-3}$, $\Omega_M=0.27$, and $\Omega_\Lambda=0.73$. 

\section{Delay Time Distributions from Volumetric SN~Ia Rates and the Cosmic Star Formation History}\label{sec:rates}

A compilation of all published SN Ia volumetric rate measurements is plotted in Figure~\ref{fig:sn1a_rates} (input data and references are in Table~\ref{tab:sn1a_rates} in Appendix~\ref{sec:ratemeas} ). For illustration purposes only, here and throughout this manuscript, we bin the rate data into 8 quantiles of nearly equal number of measures and present those binned measures, weighted by reported statistical uncertainties only, in Figure~\ref{fig:sn1a_rates} and in Table~\ref{tab:sn1a_bin}. None of the analysis presented herein was performed directly on the binned measures, rather on the individual rate values themselves.

\begin{table}[h]
   \centering
   \caption{Binned volumetric SN~Ia rates, with statistical uncertainties.}
   \begin{tabular}{lcc} 
   \hline
   \hline
   Redshift & $R_{\rm Ia}$\tablenotemark{a}&$N_{\rm measures}$\\
   \hline
$0.07 \pm{0.06}$ & $0.28^{+0.04}_{-0.03}$& 7\\
$0.19 \pm{0.06}$ & $0.30^{+0.02}_{-0.02}$& 6\\
$0.33 \pm{0.08}$ & $0.38^{+0.02}_{-0.02}$& 8\\
$0.44 \pm{0.03}$ & $0.35^{+0.05}_{-0.04}$& 6\\
$0.61 \pm{0.14}$ & $0.47^{+0.03}_{-0.03}$& 9\\
$0.81 \pm{0.07}$ & $0.60^{+0.04}_{-0.04}$& 7\\
$1.05 \pm{0.17}$ & $0.76^{+0.06}_{-0.06}$& 7\\
$1.73 \pm{0.52}$ & $0.61^{+0.14}_{-0.10}$& 7\\
\hline
   \end{tabular}
   \tablenotetext{a}{In units of $10^{-4}$ yr$^{-1}$ Mpc$^{-3}$ $h_{70}^3$}
   \label{tab:sn1a_bin}
\end{table}

A conveniently simple empirical model for volumetric rate evolution with redshift is a broken power-law evolution with redshift, $R_{\rm Ia}=R_0\,(1+z)^A$. As shown in Figure~\ref{fig:sn1a_rates}, fixing the redshift break at $z=1$ (arbitrarily) and performing a least-squares fit would give a power-law slope at $z<1$ as $A=1.55\pm0.02$ (with $R_0 = 2.40\pm0.02\times10^{-5}$ yr$^{-1}$ Mpc$^{-3}$ $h_{70}^3$), which then flattens substantially to $A=-0.1\pm0.2$ at redshifts greater than 1. This is broadly consistent with the power-law fit from \cite{Okumura:2014}, especially to $z\la1$, and consistent with recent results from the Palomar Transient Factory~\citep{Frohmaier:2019mb}. The locus is also consistent with the volumetric SN~Ia rate at $z\approx0$ converted by \cite{Li:2011b} to $2.7\pm0.3\times10^{-5}$ yr$^{-1}$ Mpc$^{-3}$ $h_{70}^3$. While the broken power-law model is useful for predicting yields from volumetric surveys, e.g., for the Wide Field InfraRed Survey Telescope \cite[\textit{WFIRST, }][]{Hounsell:2018fv} and the Large Synoptics Survey Telescope~\cite[\textit{LSST, }][]{Kessler:2019fr}, it does not inherently reveal much on the nature of SN~Ia progenitor mechanisms, which is better done through an assessment of delay-time distributions.

For these types of analyses, the standard assumption is that the stellar death rate (or supernova rate) is related to the stellar birth rate, convolved with some DTD that contains all the temporal factors of stellar evolution (e.g., main sequence lifetime, etc.) and binary star evolution (e.g., accretion rates or merger times). Two additional factors could include the fraction of the initial mass function (or IMF) that the progenitors of SNe~Ia arise from, presumably $3 - 8\, \rm{M}_{\odot}$ zero-age main sequence stars (as discussed in Section~\ref{sec:wds}) and the fraction of that population that is actually capable of producing events, as not all progenitor stars are necessarily in binary systems, or presumably the right type of binary systems to successfully result in SNe~Ia.

We can relate volumetric SN~Ia rate history to the cosmic star formation history ($\dot{\rho}_{\star}$) in a similar way, expressed mathematically by, 
\begin{equation}
R_{\rm Ia}(t) =  h^2\,k\,\varepsilon\,\biggl[\dot{\rho}_{\star}(t) \ast \Phi(t)\biggr],
\label{eqn:std}
\end{equation}
\noindent where $\Phi(t)$ is the DTD, $k$ is the fraction of the IMF (by mass) responsible for SN~Ia progenitors, $\varepsilon$ is the fraction of that population that is ultimately successful in producing SNe~Ia, and $t$ is the forward-moving clock of the universe. The two factors of the dimensionless Hubble constant ($h$) arise from the determination of stellar mass formation from luminosity in $\dot{\rho}_{\star}(t)$~\citep[see][]{Croton:2013ty}.

\subsection{The Fraction of Stars Responsible for SNe~Ia}~\label{sec:wds}
Dissecting each of the terms in Equation~\ref{eqn:std}, $k$ is perhaps the easiest to approximate. The progenitors of SNe~Ia have traditionally been CO WD which acquire sufficient mass to approach or exceed the Chandrasekhar mass limit, $M_{ch}=1.44\,M_{\odot}$. To only marginally achieve this, they can either start at sufficiently high mass to require only a small amount of accretion from a nearby companion (typically single-degenerate, or SD, scenarios), or as a pair of WD that have combined in mass to meet this criterion (the double-degenerate, or DD, scenario) setting an even lower initial mass threshold for each WD progenitor~\cite[see][ for a review]{Maoz:2014fj}. 

{In the case of DD mergers, WD mass distributions are strongly peaked around $M_{\rm WD}\approx0.6\pm0.1\,M_{\odot}$ but skewed with a significant tail extending to $1.4\, M_{\odot}$ \citep{Catalan:2008il}. A pair of WDs drawn from such distribution would be on average approximately $0.7\, M_{\odot}$ each, and together satisfactorily close to the minimum ignition threshold of a carbon core for a non-rotating CO WD, approximately $1.38\, M_{\odot}$ \citep{Arnett:1969dw, Nomoto:1982vh,Pakmor:2013gf}. Initial-Final Mass relations \cite[e.g.,][]{Catalan:2008il,Cummings:2018oe} would correspond these to zero-age main-sequence (ZAMS) masses of approximately $3\, M_{\odot}$, but no less than approximately $2\, M_{\odot}$. }

The same Initial-Final Mass relations would suggest that a WD essentially at $M_{ch}$ would fall just below $8\, M_{\odot}$ ZAMS. Similarly, simulations show that the lowest mass in which C ignition is still possible is around $6-8 \,M_{\odot}$~\citep{Chen:2014rb,Denissenkov:2015rf}, but likely no more than $\sim11\, M_{\odot}$ \citep{Takahashi:2013jx}, above which an electron-capture-induced collapse mechanism begins, marking the onset of core-collapse supernovae. {Further, stars above this mass limit form Oxygen-Neon WDs rather than CO WDs~\citep{Doherty:2017qy}}. It is reasonable, therefore, to assume a SN~Ia progenitor mass range of about $3-8\,M_{\odot}$ ZAMS. 

{Here, it should be noted that there are several other channels by which SNe~Ia could result from WD progenitors, including sub-Chandrasekhar models for explosions $\sim1\,M_{\odot}$ that involve He accretion or mass transfer and may involve more than one detonation. Similarly, there are several individual SNe~Ia that exhibit characteristics in support of these other mechanisms, for example SNe~1999by and 2018byg as potential sub-$M_{ch}$~\citep{Blondin:2018wn,De:2019rv}. However, those examples, based on comparable characteristics in observed low-$z$ samples, represent only a fraction of all SNe~Ia. The purpose of this study is to explore the dominant channel for SN~Ia production in field galaxies across all redshifts. Different conclusions could be drawn from populations in short-lived dwarf galaxies from stellar abundances~\citep{Kirby:2019aa}.}

From a numerical assessment of these stars, assuming they fall within an IMF that is a power-law distribution by mass in this initial mass range, with $\alpha\approx-2.3$~\citep{Salpeter:1955rw,Kroupa:2001gf}, one would expect { 
\begin{equation}
k = \frac{\int\limits_{3M_{\odot}}^{8M_{\odot}} N(M)\,dM}{\int\limits_{0.1M_{\odot}}^{125M_{\odot}} M\cdot N(M)\,dM}= 0.021^{+0.014}_{-0.003}\,M_{\odot}^{-1},
\label{eqn:k}
\end{equation}}
\noindent where the error in $k$ is driven more by choices in the upper and lower value in the selected mass range of SN~Ia progenitors than by the choice in IMF model, as detailed above. {The errors shown represent the 68\% confidence region  derived from thousands of realizations Equation~\ref{eqn:k}, with integration limits drawn from the range of lower mass bounds (2 to 3.5 $M_{\odot}$) and upper mass bounds (6 to 11 $M_{\odot}$).} 

The fraction of CO WDs {in the mass range of SN~Ia progenitors} that are ultimately successful in making SNe~Ia is hard to determine, as we do not yet know the details of the progenitor mechanism or mechanisms. Estimates swing rather wildly from (perhaps) as low as 1 in 200~\citep{Breedt:2017rp} to as optimistic as 1 in 40~\citep{Maoz:2012}. There is at least strong consensus that accretion on to a CO WD is essential, but that broadly describes a wide range of very different yet plausible WD close binary scenarios, at least from a theoretical standpoint~\citep{Nelemans:2001hb,Nelemans:2001cs}. The binary fractions of WDs estimated from the ESO-VLT Supernova-Ia Progenitor Survey~\citep[SPY;][]{Napiwotzki:2007,Napiwotzki:2019ez} suggest close double WD systems have $\varepsilon_{\rm bin}\simeq0.1\pm0.02$ \citep{Maoz:2017zl}. It is not likely all of these systems successfully yield SNe~Ia as their merger rates in the Milky Way would imply event rates at least a magnitude higher than best estimates of the SN~Ia rate in our galaxy, and presumably some of these systems will form AM CVn and R~Corona Borealis stars. But at least this estimate could be treated as an upper limit on $\varepsilon$.

{As is shown in the subsequent sections, in both methods of analysis, $k$ (and for that matter, $h$) are considered fixed quantities with errors that do not factor into the estimation of other parameters. We do, however, fit specifically for $\varepsilon$ and allow it to carry with it all derived scaling uncertainties. As will also be shown, those errors are much smaller ($\approx5\%$) than the uncertainties shown in Equation~\ref{eqn:k}.}

\subsection{The Star Formation Density History}\label{sec:csfh}
\begin{figure*}[t]
   \centering
   \includegraphics[width=6.1in]{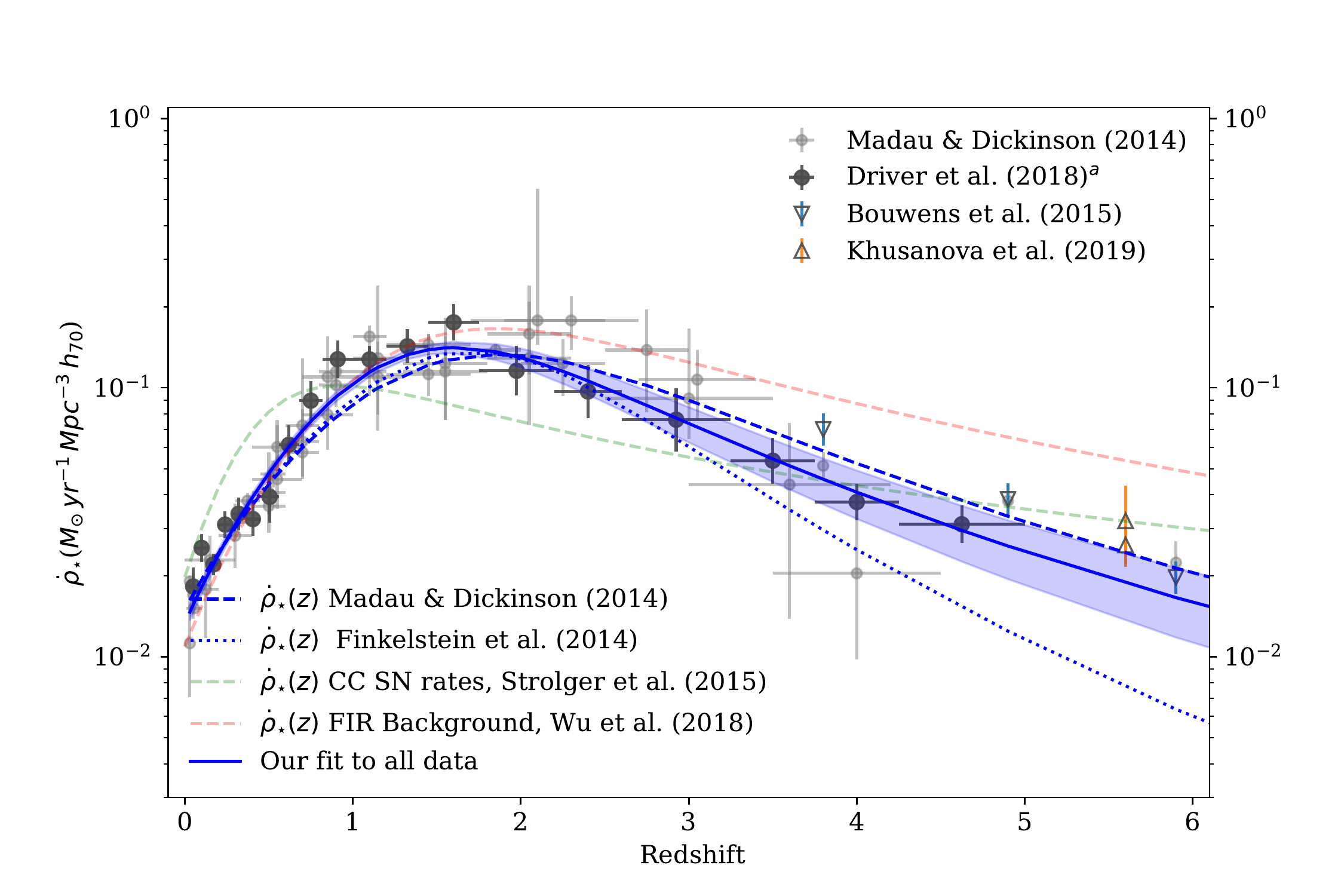}
   \caption{\footnotesize Shown are a compendium of cosmic star formation history measures, from \cite{Madau:2014fk}, \citeauthor{Driver:2018nr} (\citeyear{Driver:2018nr}; [a]-dust corrected), \cite{Bouwens:2015qy}, and \cite{Khusanova:2019kx}. Dashed and dotted lines are previous models from \cite{Madau:2014fk}, \cite{Finkelstein:2014fj}, \cite{Strolger:2015aa}, {and \cite{Wu:2018qy}}, as indicated. Solid blue line (and blue shaded region) represents our best-fit model to the compendium of data.}
   \label{fig:csfhs}
\end{figure*}

The cosmic star formation history (CSFH), at least to $z < 5$ or over 90\% of the history of the universe, is fairly well understood, with~\citeauthor{Madau:2014fk} (\citeyear{Madau:2014fk}; MD14 hereafter) providing one of the most complete compilations. More recently, the CSFH derived from the combined GAMA, G10-COSMOS, and 3D-HST datasets by~\cite{Driver:2018nr}, in a quasi-homogeneous analysis over a larger area, provides a dataset with greatly reduced uncertainties per datum, but fewer data than presented in the MD14 compendium (see Figure~\ref{fig:csfhs}).  We combine the MD14 and \cite{Driver:2018nr} data, with additional star-formation rate densities from \cite{Bouwens:2015qy} and \cite{Khusanova:2019kx}, to arrive at today's compendium CSFH using the parameterization,
\begin{equation}
\dot{\rho}_{\star}(z) = \frac{A\,(1+z)^C}{((1+z)/B)^D+1} {\rm M}_{\odot}\, {\rm yr}^{-1}\, {\rm Mpc}^{-3}\, h_{70}.\label{eqn:mdp}
\end{equation}
However to do so, we must also correct the \cite{Driver:2018nr} data for dust attenuation following the prescription in MD14, by applying 
\begin{equation}
	\dot{\rho}_{\star}(z) =h^3\,\biggl[1+10^{0.4\cdot A_{\rm FUV}(z)}\biggr]\, \dot{\rho}_{\star, {\rm uncorrected}}(z),
\end{equation}
\noindent where it is assumed $A_{\rm FUV}(z)$ has the same functional form of Equation~\ref{eqn:mdp}, with $A=1.4\pm0.1$, $B=3.5\pm0.4$, $C=0.7\pm0.2$, and $D=4.3\pm0.7$ as determined from the $A_{\rm FUV}(z)$ data from MD14. We then fit Equation~\ref{eqn:mdp} to the combined CSFH datasets, resulting in Levenberg-Marquardt least-squares solution parameters which are excellently constrained, as shown in Table~\ref{tab:csfh_fits} and Figure~\ref{fig:csfhs}.

\begin{table*}[t]
    \centering
    \caption{Cosmic Star Formation History Fit Parameter}
    \label{tab:csfh_fits}
    \begin{tabular}{lcccc}
         & $A$ & $B$ & $C$ & $D$ \\
        \hline
        \hline
	\cite{Madau:2014fk} & $0.015$ & $2.9$ & $2.7$ & $5.6$\\
	\cite{Finkelstein:2014fj} & $0.015$ & $2.9$ & $2.7$ & $7.0$\\
	\cite{Strolger:2015aa}, CCSNe& $0.015 \pm 0.001$ & $1.5 \pm 0.1$ & $5.0 \pm 0.2$ & $6.1 \pm 0.2$\\
	\cite{Wu:2018qy}, FIR background& $0.0157^{+0.0003}_{-0.0003}$ & $2.51^{+0.04}_{-0.03}$ & $3.64^{+0.04}_{-0.05}$ & $5.46^{+0.10}_{-0.09}$\\
	\hline
	\cite{Madau:2014fk}\tablenotemark{a} data fit& $0.013 \pm 0.001$ & $2.6 \pm 0.1$ & $3.2 \pm 0.2$ & $6.1 \pm 0.2$\\
	\cite{Driver:2018nr}\tablenotemark{b} data fit& $0.014 \pm 0.001$ & $2.5 \pm 0.2$ & $3.3 \pm 0.3$ & $6.2 \pm 0.3$\\
	\hline
	Combined data fit & $0.0134 \pm 0.0009$ & $2.55 \pm 0.09$ & $3.3 \pm 0.2$ & $6.1 \pm 0.2$\\
	\hline
    \end{tabular}
   \tablenotetext{a}{New fit to the cited tabular data.}\tablenotetext{b}{Corrected for dust attenuation.}
\end{table*}
 
\subsection{SN~Ia Progenitor Delay-Time Distribution Models}~\label{sec:dtd}
Theoretical DTDs result from physically constrained analyses of binary population synthesis~\cite[see][ for a review]{Wang:2012a}.  In SD scenarios, details ranging from composition of the companion donor star (H or He) to the mass-accretion efficiency lead to rather large variations in the expected DTDs~\citep{Nelemans:2013}. DD models, however, are in reasonable agreement with one another, largely because the scenario is governed by the loss of angular momentum due to the radiation of gravitational waves. The timescales involved depend on the  initial separations of the WDs. It is assumed that the population of WD binaries follow a power-law of initial radial distributions, $\Phi(r)=r^{B}$, with power $B\approx-1$ \citep{Opik:1924xr}, as is supported by SPY close WD systems, with separations distributed following $B=-1.3\pm0.2$~\citep{Maoz:2017zl, Maoz:2018eu}. It follows that the resultant delay time distribution will also follow a power-law distribution, $\Phi(t)=t^{\beta}$, with a power close to $\beta\approx-1$.

DTD recovery methods based on matching theoretical $\Phi(t)$ to CSFHs and SN~Ia volumetric rates has been hampered largely by the uncertainty in the latter two \citep{Dahlen:2008,Strolger:2010,Graur:2014,Rodney:2014fj}, specifically in the uncertainty in the SN~Ia rate above `SN high noon' (around $z\sim1$), and the uncertainty in CSFH above `cosmic high noon' (around $z\sim2$). It seems now, however, that those uncertainties have reduced to the point of making a $\Phi(t)$ reconstruction viable.

\begin{figure}[t]
   \centering
   \includegraphics[width=3.5in]{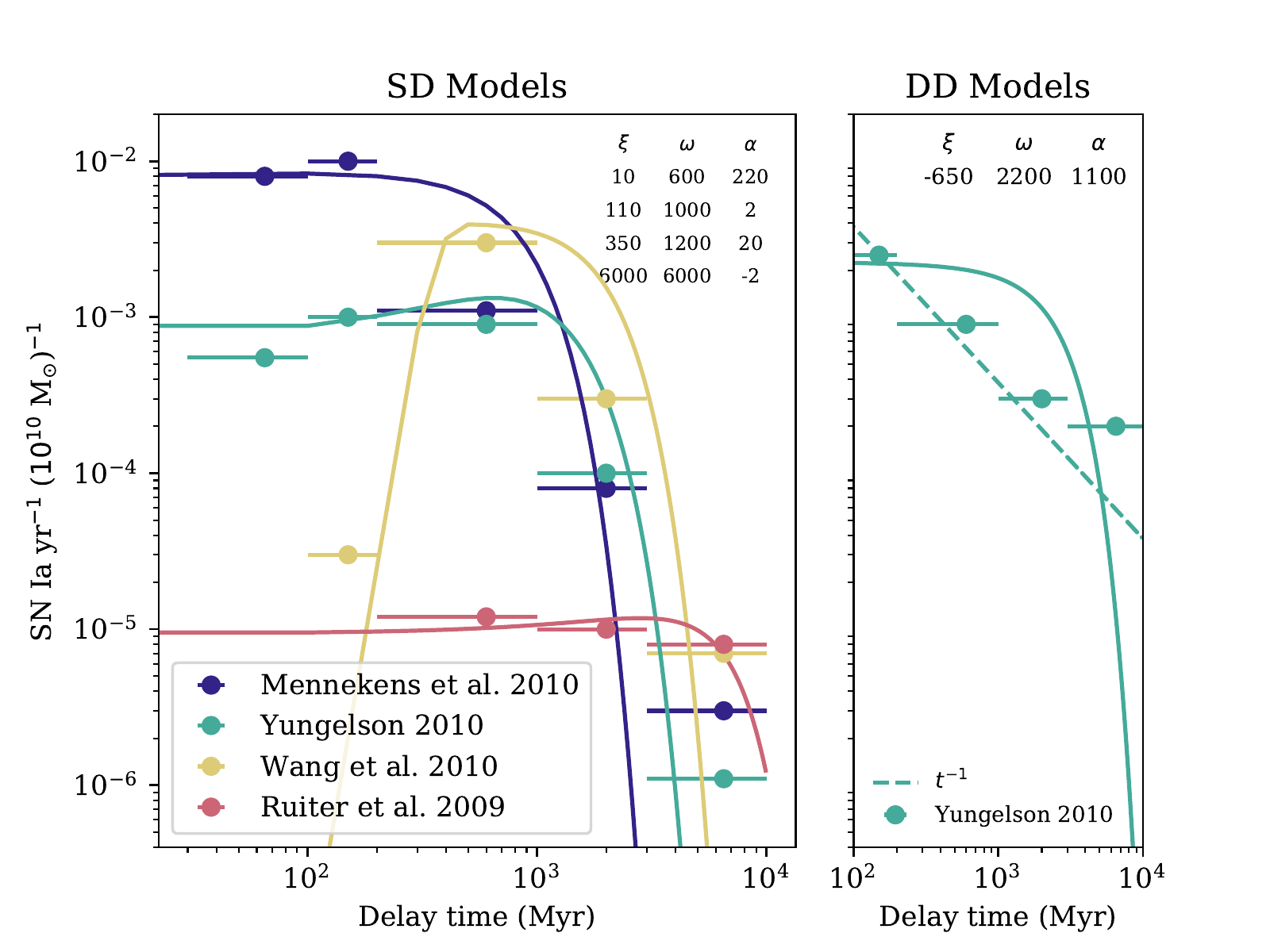}
   \caption{\footnotesize Examples of delay-time distributions from binary population synthesis analyses for SD (left) and DD (right) scenarios, from~\cite{Nelemans:2013}. Shown also (as solid line) are fits to these model bins using the function described in Equation~\ref{eqn:model}, tuning dependent variables $\xi$, $\omega$, and $\alpha$. These are the best the function can do at representing these models, if they are indeed preferred by the data.}
   \label{fig:dtd_families}
\end{figure}

Following the methodology in \cite{Strolger:2010}, we can test the intrinsic shape of the delay time distribution using a tunable unimodal model, then compare the results to the shapes of the theoretical distributions for SD and DD models. We use a skew-normal $\Phi(t)$ function, defined as:

\begin{equation}
	\Phi(t)=\frac{1}{\omega\pi}\,\exp\biggl(\frac{-(t-\xi)^2}{2\omega^2}\biggr)\int_{-\infty}^{\alpha (\frac{t-\xi}{\omega})} \exp\biggl(\frac{-t'^2}{2}\biggr)\,dt',
\label{eqn:model}
\end{equation}

\noindent where location ($\xi$), width ($\omega^2$), and shape ($\alpha$) are the dependent variables. Figure~\ref{fig:dtd_families} demonstrates the flexibility of the function in reproducing various model SD and DD from \cite{Nelemans:2013}. As can be seen, the defined function does a fairly good job of reproducing the shapes of various binary population synthesis models, particularly for SD distributions. It is also fairly reasonable in fitting DD distributions, although it should be emphasized that due to the exponential nature of the function, it has trouble \textit{exactly} reproducing the shape of a distribution that is intrinsically a power-law, a point that will be revisited in Section~\ref{sec:discussion}.

Either through an optimized fit of the functional parameters ($\xi$, $\omega$, and $\alpha$), or through a Markov-chain monte carlo (MCMC), we can test model $\Phi(t)$ through Equation~\ref{eqn:std} in comparison to the volumetric rate measurements.

\subsection{The Optimized Solution\label{sec:optimized_soln}}
We apply a maximum likelihood estimation method to determine the best-fit skew normal delay time model to Equation~\ref{eqn:std} using an optimized method described in \cite{Hogg:2010fj}. {For simplicity, we assume that the uncertainties for all published volumetric rate measurements ($\sigma_i$) are gaussian in nature, but may be underestimated by some factor ($f$) that scales with the value of the observed rates. This is motivated by the fact that we are using just the statistical error reported for each rate value, and the most plausible sources of systematic uncertainty (such as classification errors and misestimated detected efficiencies) will tend to increase as the observed rates increase.} As follows, we adopt the likelihood function to be:
\begin{eqnarray}
&\ln p (y|x, \sigma, \varepsilon, \xi, \omega, \alpha, f) = &\nonumber \\
&-\frac{1}{2} \sum_i \biggl\{ \frac{[R_{{\rm Ia},i} - R_{\rm Ia}(t_i; \varepsilon, \xi, \omega, \alpha)]^2}{s_i^2}+\ln (2\pi s_i^2)\biggr\}&,
	\label{eqn:lf}
\end{eqnarray}
where,
\begin{equation}
s_i^2 = \sigma_i^2+f^2\, R_{\rm Ia}(t_i; \varepsilon, \xi, \omega, \alpha)^2,
\end{equation}
\noindent $R_{{\rm Ia},i}$ are the various independent rate measures, and $R_{\rm Ia}(t_i)$ are the parameter-dependent model predictions at the cosmic time of the various rate measures. We then find the optimal parameters which maximize this likelihood. 

\begin{figure}[t] 
   \centering
   \includegraphics[width=3.5in]{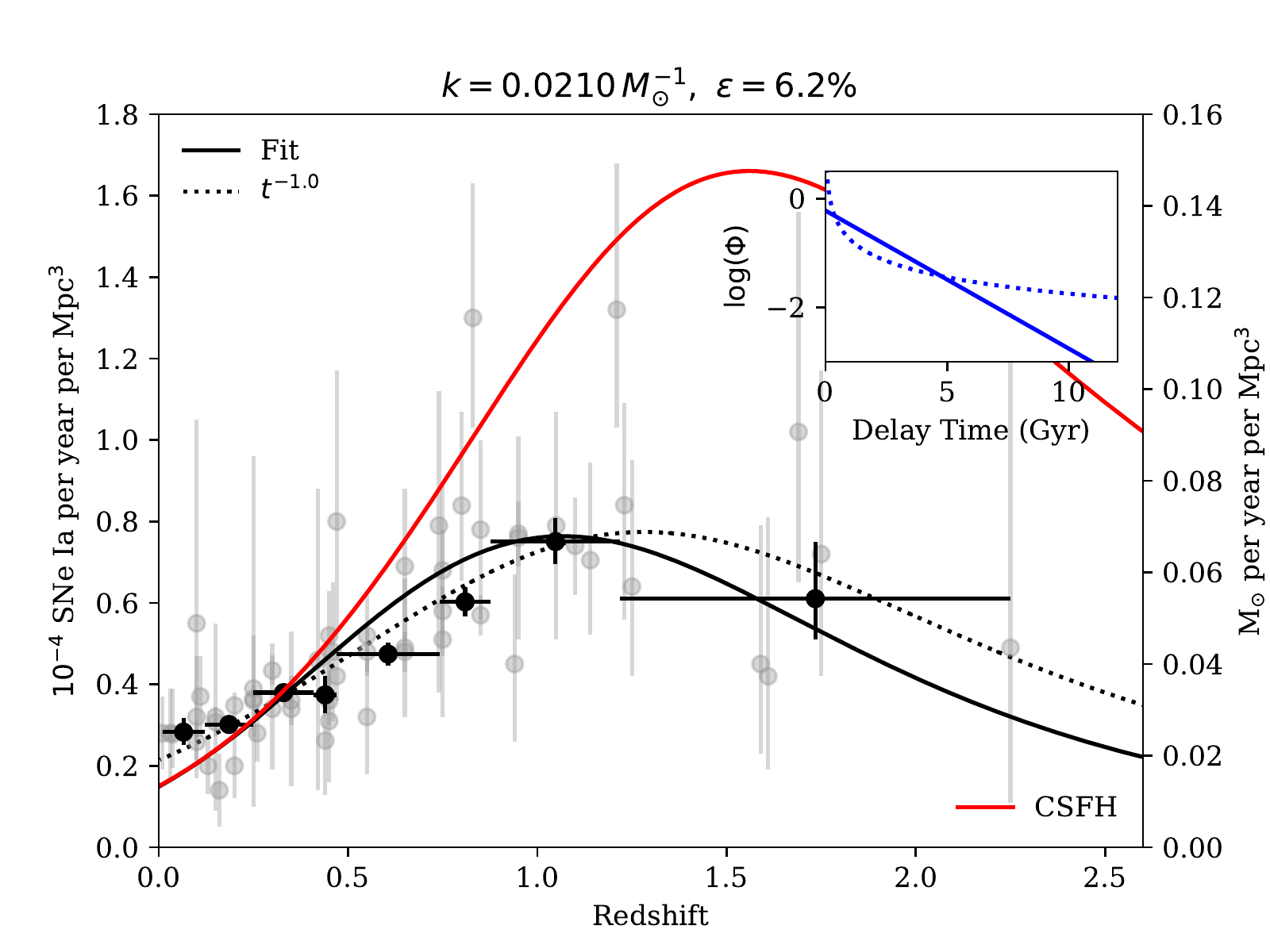} 
   \caption{\footnotesize In addition to rate values shown in previous figures, the $R_{\rm Ia}(z)$ model results from optimal parameter fitting of the unimodal $\Phi(\tau)$ model is shown (solid black line) in comparison to a $\beta=-1$ power-law $\Phi(\tau)$ (dashed black line). The inset shows the comparison of the two $\Phi(\tau)$ models. The CSFH is shown on in red, and along the secondary ordinate. }
   \label{fig:sfd_optimized_curvefit}
\end{figure}

As for priors, we require the successful fraction of progenitors to be between zero and unity ($0<\varepsilon<1$), that the width parameter can only be positive ($\omega>0$), and that the underestimation fraction can only be between approximately zero and unity ($-4<\ln f<0$). Otherwise, we apply rather loose and arbitrary bounds of $-2000<\xi<2000$ and  $-500 < \alpha < 500$. The results of this optimized fit are shown in Figure~\ref{fig:sfd_optimized_curvefit} and tabulated in Table~\ref{tab:results}.

The optimization results in a model that is seemingly consistent with the $t^{-1}$ model, although it is not directly possible to estimate errors on the best-fit parameters, or the range of validity via this maximum likelihood optimization method. 

\begin{table*}[t]
    \centering
    \caption{Results for Skew-normal Model fits}
    \label{tab:results}
    \begin{tabular}{ccccccc}
        \hline
                Model test & Sections & $\ln \varepsilon$ & $\xi$ & $\omega$ & $\alpha$ & $\ln f$ \\ 
                \hline
		CSFH Max.~Likelihood (Optimized)&\S\ref{sec:optimized_soln}& $-2.78$&$-1518$& $51$& $50$& $-2.41$\\
                CSFH MCMC &\S\ref{sec:mcmc_sfd}& $-2.81^{+0.05}_{-0.05}$ & $-1258^{+523}_{-669}$ &$59^{+18}_{-12}$& $248^{+169}_{-171}$&  $-2.6^{+0.8}_{-0.7}$\\
                SFH MCMC &\S\ref{sec:sfh}& $-2.88^{+0.14}_{-0.13}$ & $-1076^{+506}_{-624}$ &$78^{+21}_{-15}$& $226^{+157}_{-175}$&  \nodata\\
                \hline
    \end{tabular}
\end{table*}

By way of performing a more direct comparison of the quality of the models as fits to the volumetric rate data, we calculate the Akaike information criterion (AIC) and the Bayesian information criterion (BIC) for the optimized and $\beta=-1$ power-law solutions, as well as the median MCMC solutions calculated in the next sections, which are shown in Table~\ref{tab:aicbic}. {The AIC and BIC are statistical tools that effectively formalize Ockham's razor$-$ punishing models for the use of free parameters to provide flexibility that is not needed to accurately reproduce the data~\citep{Akaike:1998fp,Schwarz:1978qd}. As both tests estimate the amount of information lost, lower values indicate a higher quality fit of the model to the data. The $t^-1$ models is preferred by the AIC/BIC tests because an adequate match to the data, and does not have as much flexibility as our skew-normal models. However, one should also note that neither of these criteria take into account uncertainties in the data.}

\begin{table}[h]
	\centering
	\caption{AIC/BIC}
	\label{tab:aicbic}
	\begin{tabular}{cccc}
		\hline
		Model test & Sections & AIC & BIC\\
		\hline
		$t^{-1}$&\S\ref{sec:optimized_soln} & $3.0$ & $-202.0$\\
		CSFH Max.~Likelihood (Optimized)&\S\ref{sec:optimized_soln}& $6.9$ & $-196.3$\\
		Median CSFH MCMC&\S\ref{sec:mcmc_sfd}& $6.4$ & $-181.2$\\
		Median SFH MCMC&\S\ref{sec:sfh}& $5.5$ & $-153.4$\\
		\hline
	\end{tabular}
\end{table}

\subsection{The MCMC solution\label{sec:mcmc_sfd}}
\begin{figure}[t]
   \centering
   \includegraphics[width=3.5in]{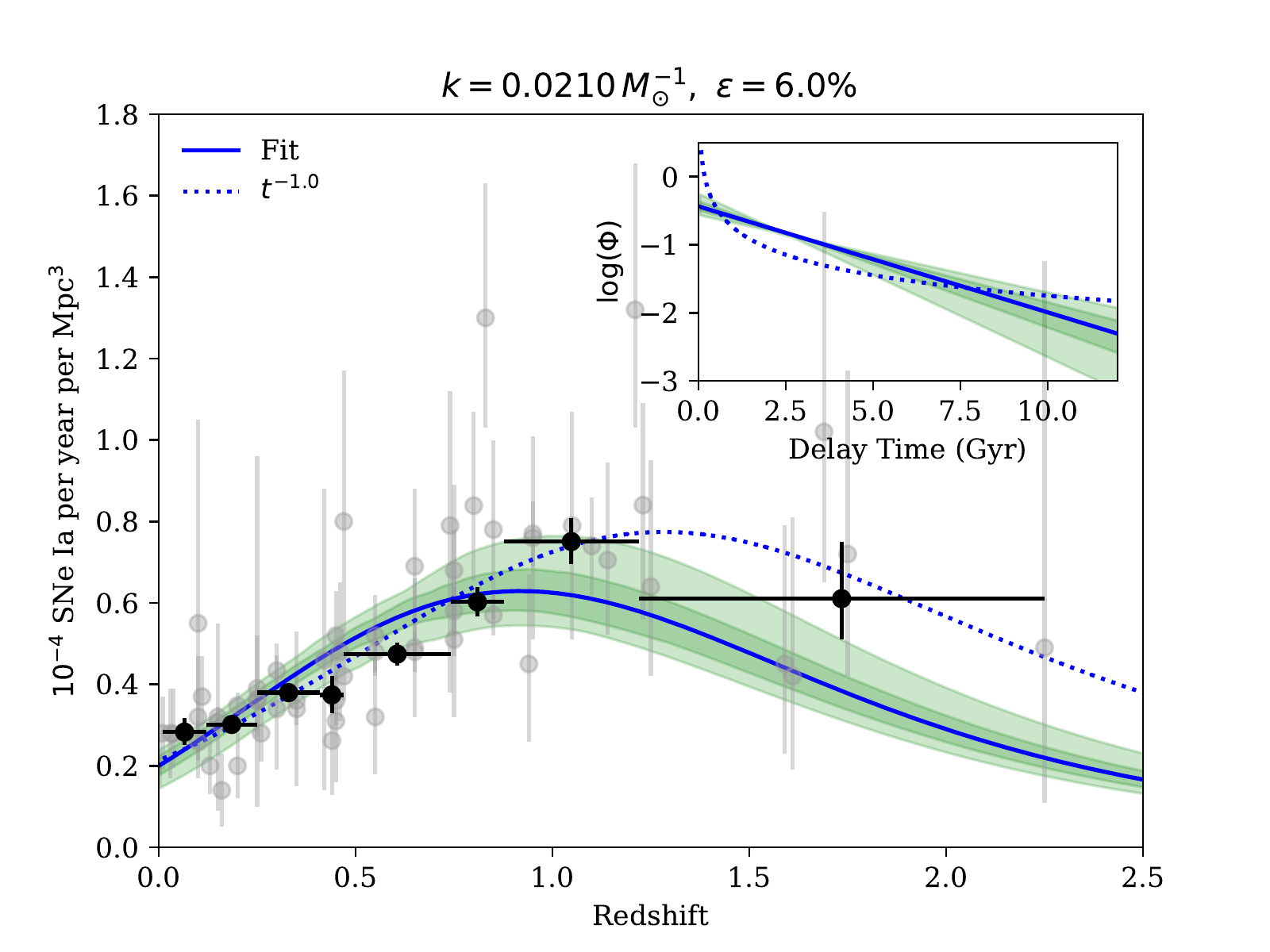} 
   \caption{\footnotesize Similar to Figure~\ref{fig:sfd_optimized_curvefit}, the $R_{\rm Ia}(z)$ result of from MCMC best-fit is shown (blue line), with the 68\% and 95\% confidence intervals, in dark and light green, respectively. }
   \label{fig:figure_fit_demo_werr}
\end{figure}

Exploring the parameter space in an MCMC allows both confirmation of the optimized solution and an exploration of the range of validity. We use the affine-invariant MCMC ensemble sampler from {\tt emcee.py}~\citep{Foreman-Mackey:2013pd} using the same likelihood function as shown in Equation~\ref{eqn:lf}, and set our uniform priors as described by the bounds, as shown in the previous section, with the exception of evaluating $\ln \varepsilon$ rather than $\varepsilon$ to allow MCMC step sizes of order unity, and using the prior $ -10 < \ln \varepsilon < 0$. We then set 1000 walkers to explore 10,000 steps, for a total of 10 million iterations, the first 100,000 of which are discarded as `burn-in'. The MCMC likelihood distributions are presented in the Appendix~\ref{sec:mcmc_results}. The median solution and confidence range is shown in Figure~\ref{fig:figure_fit_demo_werr} and tabulated in Table~\ref{tab:results}.

As these results show, there is a clear convergence in $f$, the factor by which statistical errors in rate measures are collectively misestimated. We find that $\ln(f) =-2.8^{+0.6}_{-0.8}$, which means that the statistical uncertaintainties are collectively underestimated, implying a bulk systematic uncertainty in the range of $\sim4\%$ to 17\%. So, while there is a large dispersion in rate values, these values are reasonably consistent to within statistical errors, which themselves are not grossly misestimated. The fraction $\varepsilon$ is also very well constrained, with only $6.0\pm5\%$ of WD stars contributing as SN~Ia progenitors. The Hubble-time integrated SN Ia production efficiency, i.e., combining the $k$ and $\varepsilon$ terms, yields $N/M_{\star}=1.26^{+0.83}_{-0.18}$ events per 1000 M$_{\odot}$ formed (see Appendix~\ref{sec:snum} for a discussion on mass-weighted SN rate histories).

However, the parameters which set the shape of the delay time distribution,  $\xi$, $\omega$, and $\alpha$, appear very much less constrained by the MCMC. There is a clear maximum at $\omega\approx60$ that is also highly degenerate with the value of $\xi$.  And there does not appear to be any convergence or preference in the value of $\alpha$. While it would appear there is no specific solution to the function preferred by the data, the resultant range in parameters indicate a family of solutions that are indeed related. Characterized by highly negative locations and broad widths, the only part of the distributions (in the 99\% confidence interval) which lie in the positive-time domain are the exponential-like tails, as is shown in the inset of Figure~\ref{fig:figure_fit_demo_werr}, a point to be further discussed in Section~\ref{sec:discussion}. {It should be noted that the goal of this study is more model fitting than parameter estimation, utilizing a proxy function that has the flexibility to encompass the range in real delay-time distributions derived in binary synthesis models.}

\section{Delay Time Distributions from Star Formation Histories}\label{sec:sfh}
\cite{Maoz:2011} detailed a prescription for recovering delay-time distributions from an analysis of the star-formation histories of individual galaxies, both those which host SNe~Ia and those that do not, in the duration of a continuous survey. Here, we present an evaluation of the maximum likelihood delay time distribution following the same approach, but performed on the GOODS/CANDELS SN~Ia hosts and other field galaxies. 

For this analysis we use the star formation histories (SFHs) for galaxies in the GOODS/CANDELS survey area, derived using the Bayesian modeling approach of \cite{Pacifici:2012ve}. In summary, the galaxy physical properties are retrieved from a combined analysis of stellar and nebular emission utilizing an extensive library of star formation and chemical enrichment histories. These libraries build a large repository of rest-frame galaxy spectral energy distributions, which are then used to determine likelihood distributions of physical parameters from a Bayesian analysis of observed spectral energy distributions. This method has been applied to the HST/WFC3-F160W-selected CANDELS photometric catalogs for the GOODS-South \citep{Guo:2013rp}, and the GOODS-North \citep{Barro:2019vn}, and converted to SFH catalogs~\cite[see][]{Pacifici:2016ul}. For simplicity in this analysis, we adopt only the median derived SFH of each galaxy.

For a given galaxy, the rate history of SNe~Ia per year ($r_i$) would be expressed as:
\begin{equation}
r_i (t) = h^2\,k\,\varepsilon\, \int_0^t \Psi_i(t')\,\Phi(t-t')\,dt',
\label{eqn:rate_history}
\end{equation}
\noindent where $\Psi_i$ is the SFH of the galaxy (mapped in look-forward time), and $\Phi$ is the global DTD model, also in look-forward time. The product of the rate at the observed epoch ($r_i$) and the observed control time ($t'_{c, i}$) for the galaxy-- which contains all the information on the temporal sampling and depth of the survey--  give the expected number of observed SN~Ia events ($m_i$) over the duration of the survey, by
\begin{equation}
m_i = r_i \, t'_{c, i}.
\end{equation}
\noindent The probability distribution for those observed events follows a Poisson distribution, where the likelihood of catching $n_i$ SNe~Ia from a galaxy when $m_i$ are expected is
\begin{equation}
P(n_i | m_i) = \frac{m_i^{n_i}e^{-m_i}}{n_i!}.
\end{equation}
\noindent The product of probabilities for all galaxies in the survey would then serve as the likelihood of a given rate model, tuned by the chosen DTD model. The log-likelihood, which is convenient for MCMCs, is then expressed by:
\begin{equation}
L = \prod _i^N P(n_i|M_i) \Rightarrow \ln L = -\sum^N m_i+\sum^N\ln\biggl(\frac{m_i^{n_i}}{n_i!}\biggr)
\end{equation}
\noindent in which the last term is zero for the galaxies which did not host SNe~Ia during the survey.

\begin{figure}[t] 
   \includegraphics[width=3.5in]{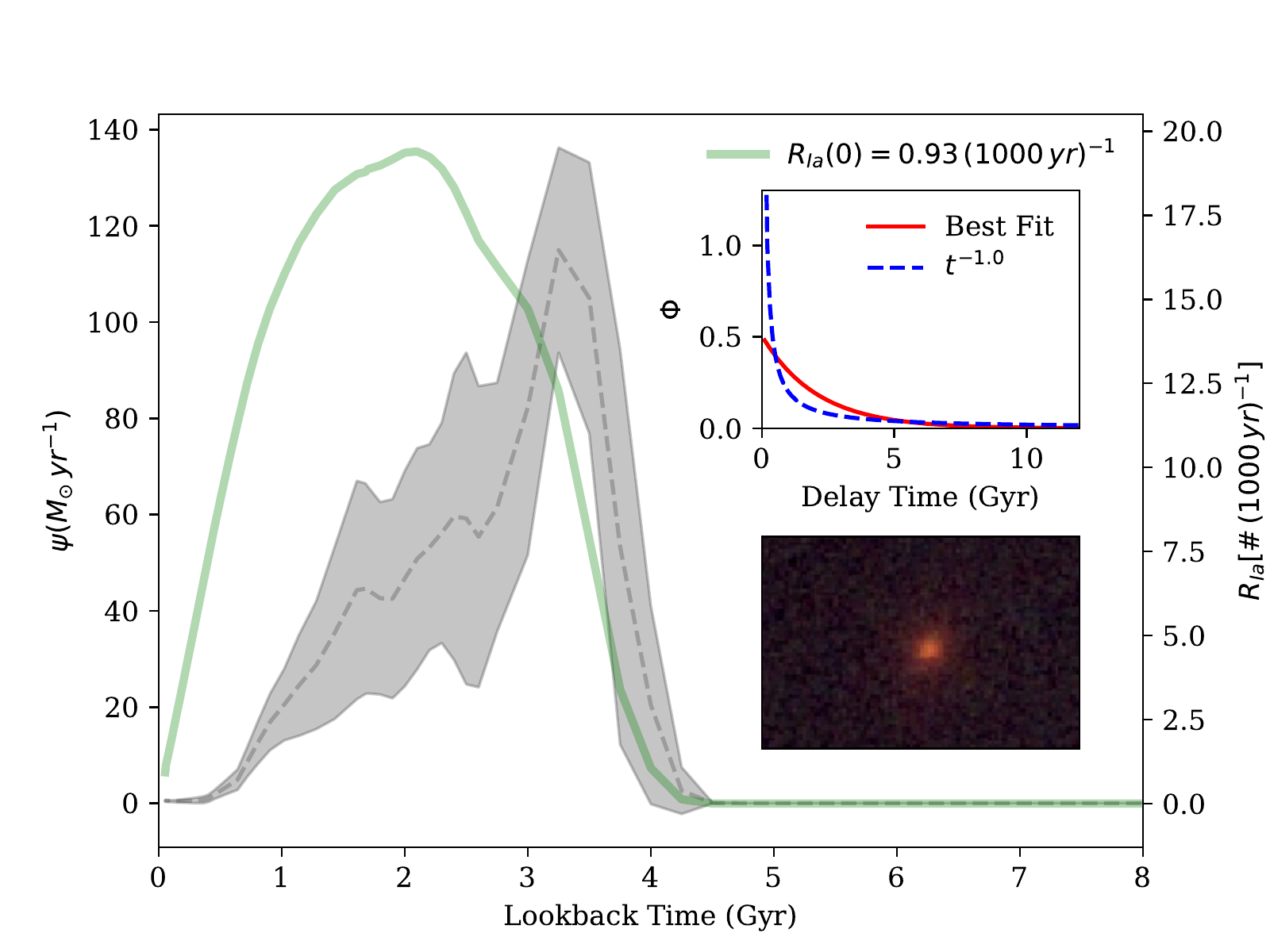} 
   \includegraphics[width=3.5in]{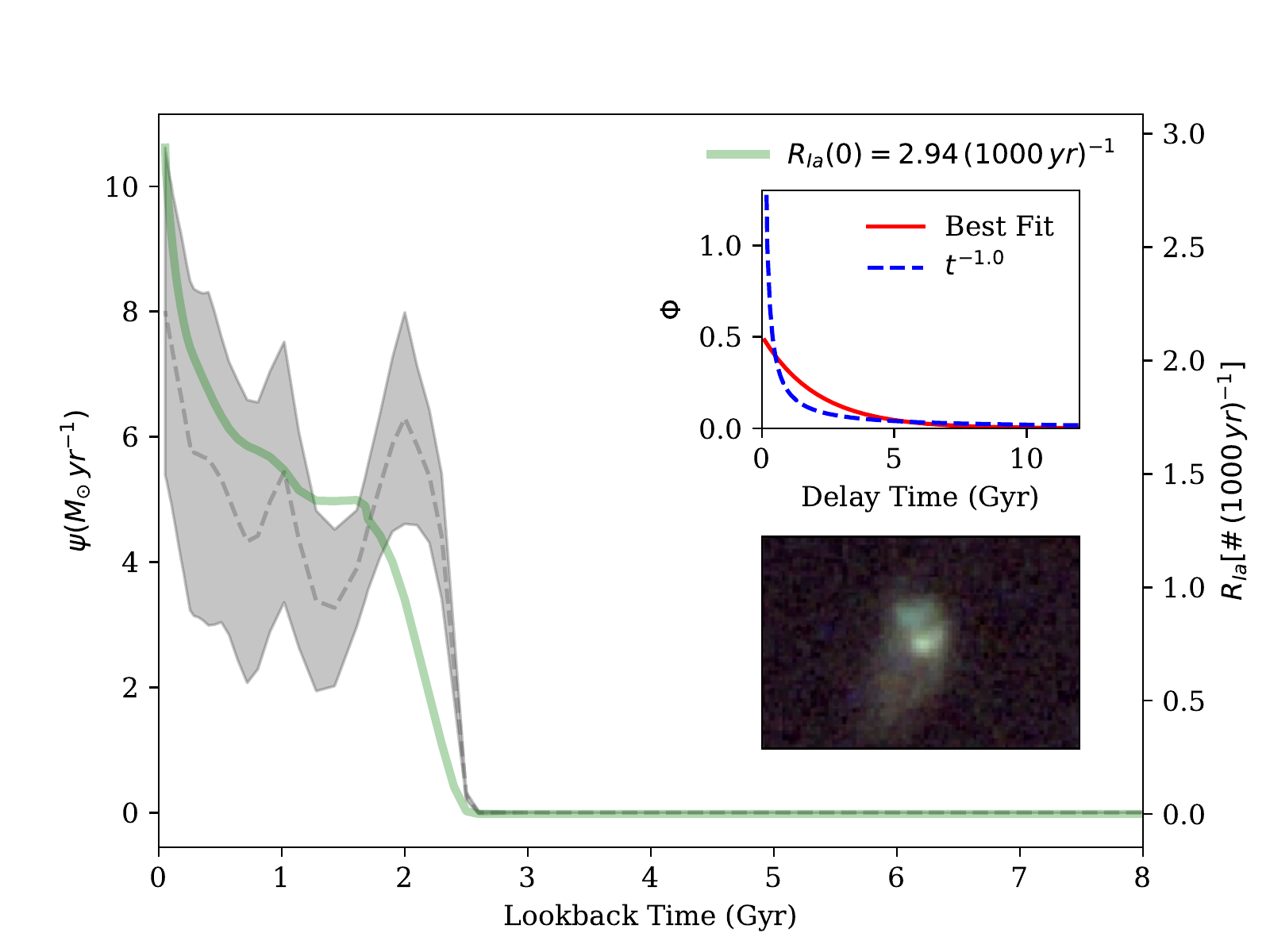} 
   \caption{\footnotesize Example star formation histories (dashed-line with gray uncertainty regions, and left ordinate), and the resultant SN Ia rate histories (green solid line and right ordinate) for two SN Ia host galaxies in our sample, SNe 2002hp (left) and 2003dy (right), in the GOODS-South and GOODS-North fields, respectively. Insets show the delay time distribution applied (upper right, in red) compared to $t^{-1}$ (blue dashed), and a three-color HST ACS/WFC image (lower right) of the SN host galaxy.}
   \label{fig:figure_sfh_fit_demo}
\end{figure}

Using the control times derived for each survey field using the methods described in~\cite{Strolger:2015aa}, Figure~\ref{fig:figure_sfh_fit_demo} shows example ``SN~Ia rate histories'' one would derive from Equation~\ref{eqn:rate_history} using the median-value models from the MCMC on CSFHs done in the previous section~{(uncertainties on the derived SFHs are shown in the shaded regions)}. The figure shows star-formation histories for two SN Ia host galaxies, for SN 2002hp and SN 2003dy, respectively \cite[see][ for further details on these events]{Strolger:2004}. Both galaxies are at $z\approx 1.3$, and in the GOODS-South and GOODS-North fields, respectively. The host of SN~2002hp is a fast-forming/slow-quenching passive galaxy that underwent a very large burst of star formation just a few Gyr ago. When that SFH is convolved with the DTD, it results in a relatively large expected rate of 0.93 SNe Ia per millennium at the observed epoch.  Conversely, the host of SN~2003dy is actively star forming at the observed epoch, albeit at a more modest rate, and has been active over the last few Gyr. The convolved result is a SN Ia rate about three times larger than the host of SN~2002hp, 2.94 events per millennium at the observed epoch. Nearly all non-hosts have predicted SN Ia rates at their respective observed epochs several orders of magnitude smaller than these two example hosts with this assumed $\Phi(t)$.

\begin{figure}[t] 
   \centering
   \includegraphics[width=4in]{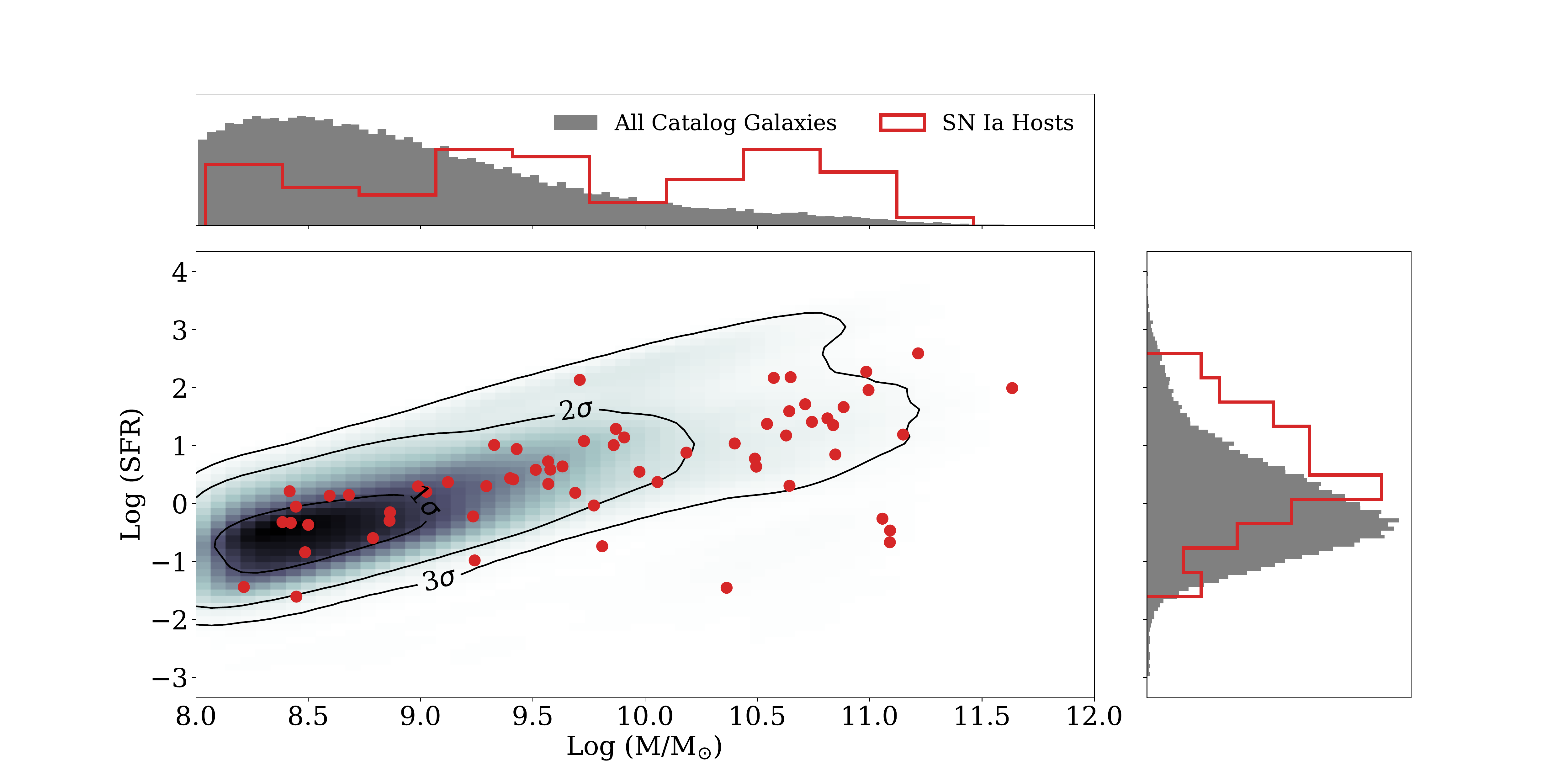} 
   \caption{\footnotesize {Joint distribution in total masses and star formation rates for the 67 SN~Ia hosts (in red) in comparison to the 68,931 GOODS-North and South catalog galaxies (gray). The SN~Ia hosts are fairly evenly spread through the range of catalog properties.}}
   \label{fig:figure_cdfs}
\end{figure}

The SFH catalog contains 70,375 {$H$-band selected sources, of which 1,444 have SExtractor~\citep{Bertin:1996,Bertin:yg} CLASS\_STAR $> 0.8$ and are deemed most likely stars, leaving a remainder of 68,931} galaxies in the GOODS-North and South fields. There are 34 events classified as SNe~Ia in the GOODS-South field, and 39 in the North field \citep{Strolger:2004, Dahlen:2008, Rodney:2014fj}, all but 6 of which were matched to host galaxies in the SFH catalog. Two of these hosts were rejected as the spectroscopic redshifts of the associated SNe were very inconsistent with the SFH catalog redshifts, and the other 4 were simply not matched to galaxies in the SFH catalog, as they were either too faint or too near the field edge to be listed in the composite photometry catalogs. {The distribution of the remaining 67 SN~Ia hosts in total masses and star formation rates (for the observed epoch), relative to the population of catalog galaxies are shown in Figure~\ref{fig:figure_cdfs}. While it cannot be said that the host population is representative of the catalog, the hosts do adequately cover the range in mass and star formation rate of the catalog, and are not biased to some extrema. The SN~Ia host do tend to be more massive and more actively star forming than the galaxy population as a whole.}

{The model parameters, $\varepsilon$, $\xi$, $\omega$, and $\alpha$ were then explored via {\tt emcee.py} in a method similar to what was done in Section~\ref{sec:mcmc_sfd}, keeping the same uniform priors as bounds. We set 100 walkers exploring 225 steps on these parameters, the first 50 of each discarded as burn-in. The maximum likelihood results for the 17,500 iterations on SFHs are identical to the results for the CSFH assessment presented in Section~\ref{sec:mcmc_sfd}, as is shown in Table~\ref{tab:results}. The MCMC likelihood distributions are presented in the Appendix~\ref{sec:mcmc_results}.}

\section{discussion}\label{sec:discussion}
\begin{figure}[t]
   \centering
   \includegraphics[width=3.5in]{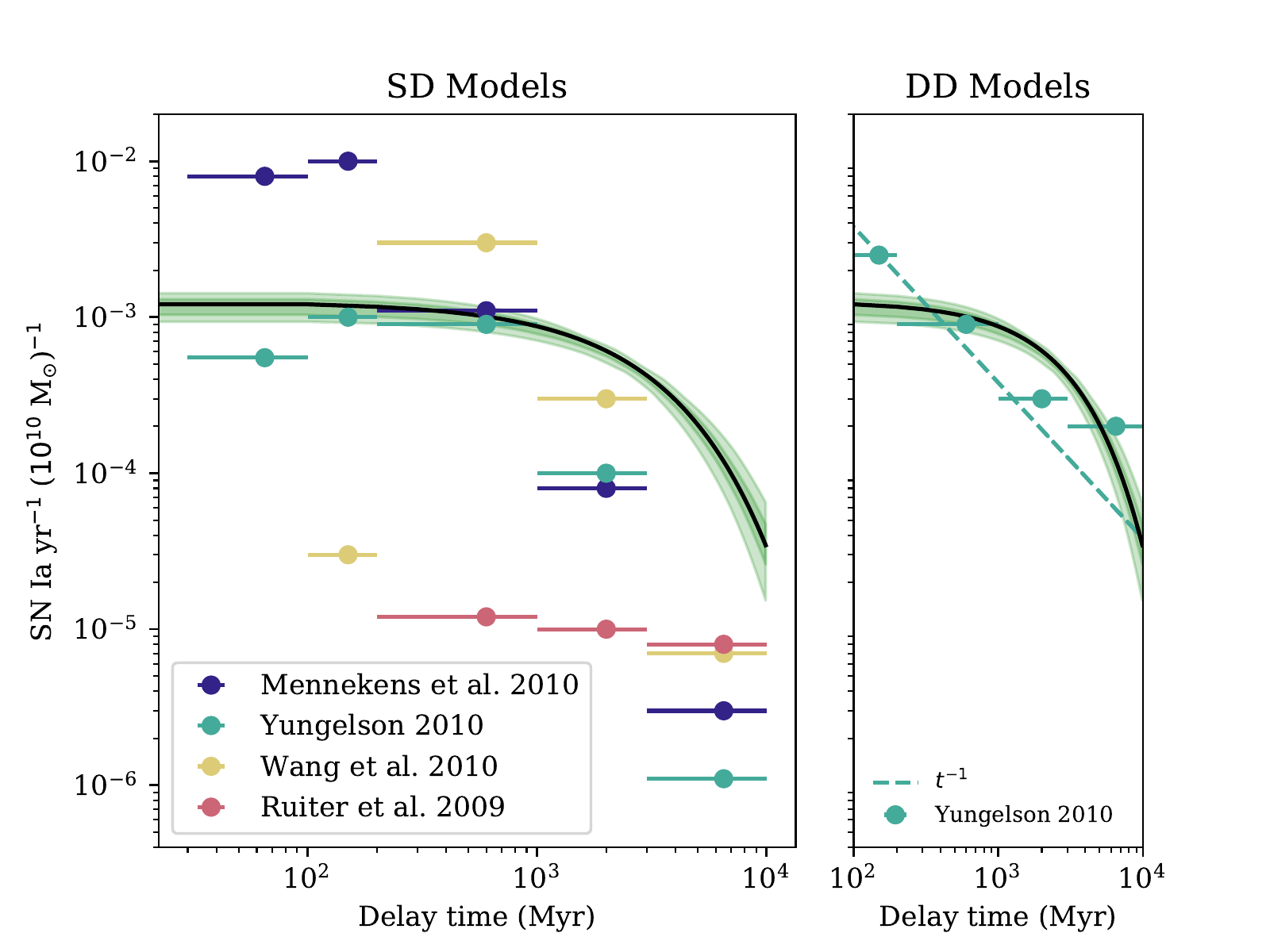}
   \caption{\footnotesize Similar to Figure~\ref{fig:dtd_families}, but showing the maximum likelihood model fit.}
   \label{fig:dtd_eval}
\end{figure}

The results of the analyses in Sections~\ref{sec:rates} and \ref{sec:sfh} point to a family of delay-time distribution models that are essentially exponential in shape, having fewer prompt events in the 40 Myr to 1 Gyr range than are expected in the $>1$ Gyr range. These model shapes are generally inconsistent with the results from SD binary population synthesis models, as is shown in Figure~\ref{fig:dtd_eval}. They also seem qualitatively consistent with DD binary population synthesis models, with the caveat that the exponential functional form of our model has difficulty reproducing the distributions that are inherently power-law, as previously indicated. Yet, despite this analytical limitation, it is worth further exploring the comparison in the exponential result from this analysis with the assumed power-law distributions that are generally in favor by the community.

\begin{figure*}[t] 
   \centering
   \includegraphics[width=6.1in]{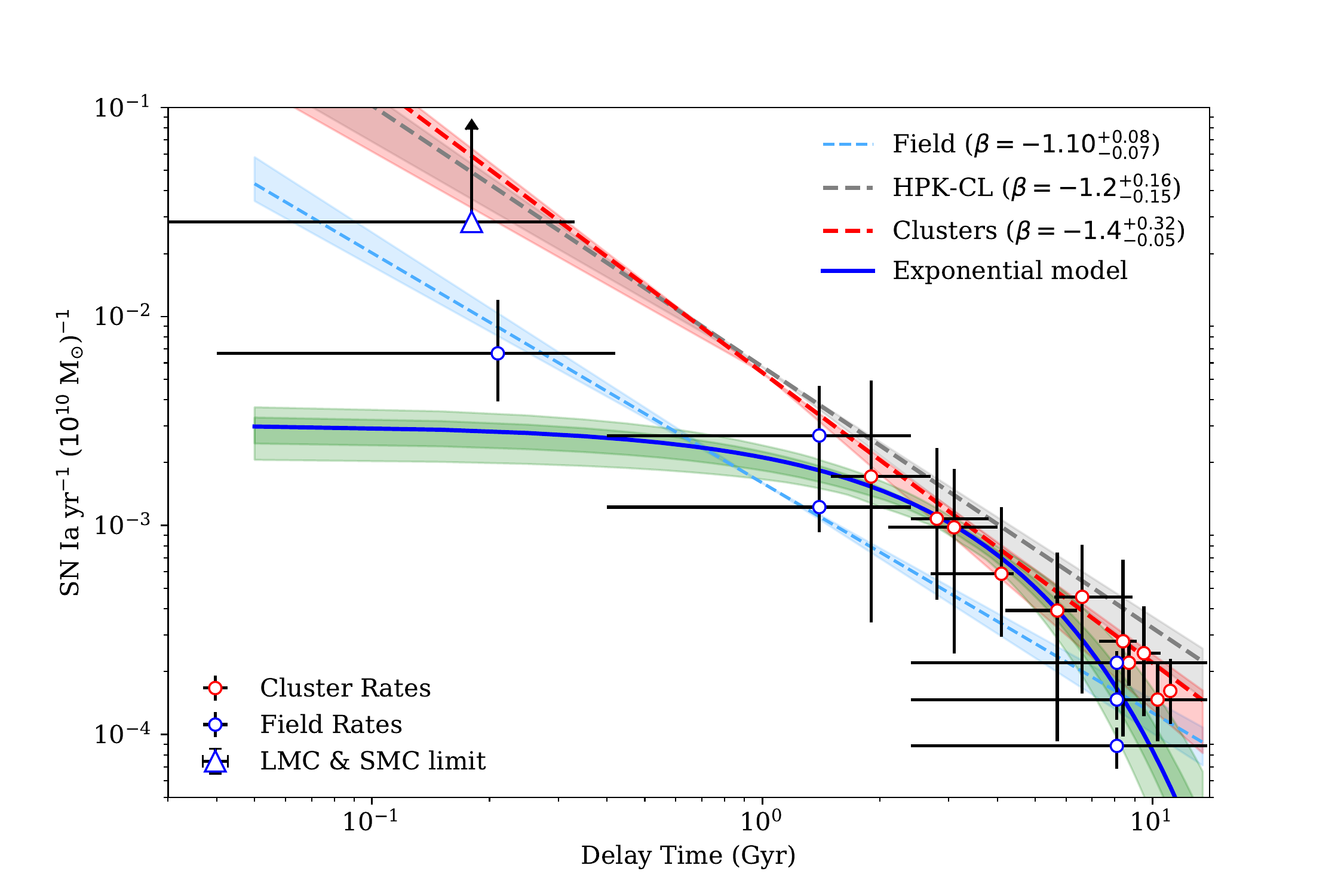}
   \caption{\footnotesize Reconstructed delay time distributions from various authors, in the LMC \& SMC, and in field and galaxy cluster environments. Shown also are the power-law model fits, and associated error regions, for field (in light blue) and cluster (in red) environments{, from similar SFH investigations. The power-law fit from an analysis of field galaxies using a relationship between sSFR and $g-r$ color is also shown (in gray) for comparison~\citep[HPK-CL,][]{Heringer:2019ws}. }Overplotted is the exponential DTD model (solid blue line) and error region (in green) derived from the analyses in Sections~\ref{sec:rates} and \ref{sec:sfh}. }
   \label{fig:logsnrate}
\end{figure*}

Figure~\ref{fig:logsnrate} shows the DTDs recovered in various time bins from SFH analyses of the Magellanic Clouds \citep{Maoz:2010}, of Sloan galaxy data~\citep{Maoz:2010a, Maoz:2011, Maoz:2012a,Graur:2013}, and from high$-z$ cluster rates~\citep{Friedmann:2018hq}, along with the derived  slopes for power-law models for SN~Ia hosts in field ($\beta=-1.1^{+0.08}_{-0.07}$) and galaxy cluster ($\beta=-1.4^{+0.32}_{-0.05}$) environments {(scaled by 1.6 and 5.4$\times10^{-12}$ M$_{\odot}$ yr$^{-1}$, respectively)}. Also shown are the exponential best fit models from our analysis, overplotted on the field and cluster data. As a relative goodness-of-fit test, we find  $\chi^2_{\nu}=0.21$ for cluster hosts ($\nu=10$) relative to $\beta=-1.3$, $\chi^2_{\nu}=3.7$ for field hosts ({$\nu=5$}, excluding the LMC+SMC upper limit) relative to $\beta=-1.1$, and $\chi^2_{\nu}=1.7$ ($\nu=17$) for all data compared to our best fit model (also excluding the LMC+SMC upper limit). It appears our exponential model is just as good as the power-law models at describing these recovered delay-time measurements, and has the added benefit of not having to invoke different slopes (or presumably different progenitor channels) for field and clustered SN~Ia host environments. {It should be noted that \cite{Heringer:2019ws} show a method for arriving at the DTD power-law slope using a relation between the specific supernova rate (sSFR)\footnote{Star-formation rate over the total stellar mass at the observed epoch.}  per unit luminosity and the color ($g-r$) of a given galaxy~\citep{Heringer:2017fp}, resulting in $\beta=-1.25^{+0.16}_{-0.15}$ (scaled by $5.8\pm1.3\times10^{-12}$ M$_{\odot}$) for the selection of galaxies from the SDSS DR7 Stripe 82. The result is higher than the reconstructed delay-time values for field galaxies, but somewhat consistent with the values in clusters of galaxies with $\chi^2_{\nu}=2.2$ ($\nu=10$). }

Another common method to testing progenitor models is by comparing the measured rate of SNe~Ia in high specific sSFR galaxies to that in their low-sSFR counterparts, in the modern `A+B'-model tests~\citep{Scannapieco:2005,Smith:2012lr,Gao:2013kk,Andersen:2018dp}. While this grossly addresses the promptness of some fraction of SNe~Ia, the test is inherently limited as it incorrectly assumes the observed SN~Ia rate is directly tied to the (A) total mass and (B) the current rate of star formation in a host galaxy, rather than appropriately connecting that SN rate to some past epoch of star formation. That, and the large uncertainties in SN rates that result from complex star-formation rate histories, are the largest sources of error in these tests.

\begin{figure*}[t]
   \centering
   \includegraphics[width=6.1in]{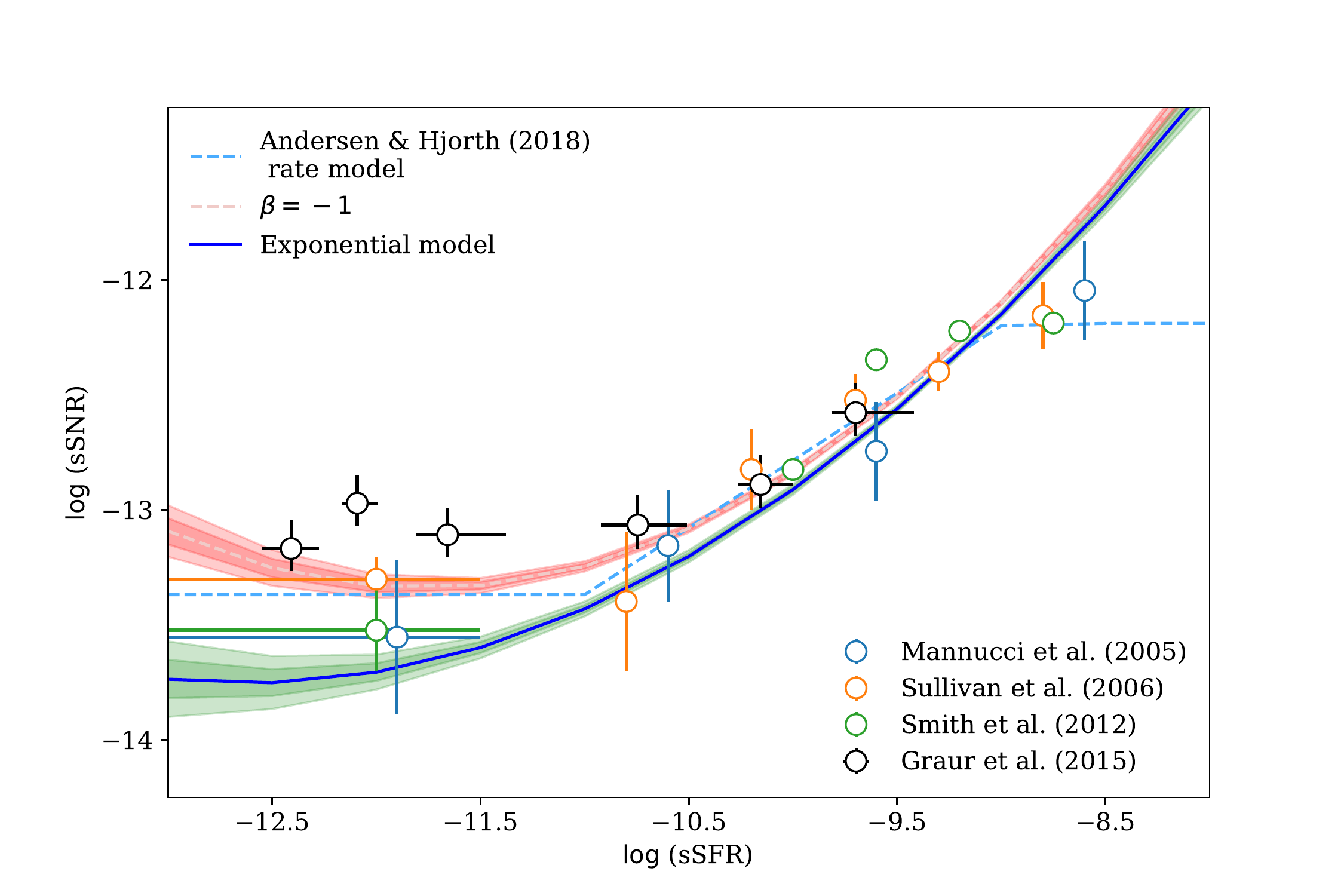}
   \caption{\footnotesize Measurements of the specific SN~Ia rate relative to specific star formation rates of their hosts, from various authors. Also shown are the expected track (and associated error region) for SNe~Ia from this analysis (blue/green region), from $\beta\approx-1.1$ power-law DTDs (in red), and a piecewise model~\cite[][ blue-dashed line]{Andersen:2018dp}.}
   \label{fig:ssfr}
\end{figure*}

Using the methods described in Section~\ref{sec:sfh} to convolve, for each SN~Ia host, the recovered SFH by the best derived DTD to get the SN rate at the observed epoch, and using catalog SFR and total masses at the observed epoch, we derive a track (and associated uncertainty region) in which SNe~Ia should lie in specific SN rates (sSNR) as a function of sSFR, as shown in Figure~\ref{fig:ssfr}. This method is similar to, but inherently more direct than that done in \cite{Graur:2015fk}, where in the latter the SFHs are estimated by an exponential law, following \cite{Gallazzi:2005rf} and \cite{Kauffmann:2003sj}, and the references therein. This is shown in comparison to measurements from \cite{Mannucci:2005}, \cite{Sullivan:2006a}, and \cite{Smith:2012lr}, and in comparison to tracks expected from $\beta\approx-1.1$ power-law delay time models, and a piecewise model from \cite{Andersen:2018dp}. The measurements are consistent with all three tracks~{in the region for star-forming and `burst' galaxies. However, the tracks strongly diverge in the region of passive galaxies, where sSFR $\la10^{-11}$ M$_{\odot}$ yr$^{-1}$, and where the measurements show their highest scatter}. It is expected that further studies in passive galaxies will provide some clarity. Additionally, tests of these tracks in field dwarf galaxies, specifically those not associated with clusters of galaxies, may be illuminating as they are simpler to model by virtue of having many fewer episodes of star formation. 

As a final note, now that the evidence for double white dwarf mergers as the primary source of SNe~Ia reaches concordance, it is increasingly interesting to further investigate the fine details of the exact progenitor mechanism. 

\begin{figure}[t] 
   \centering
   \includegraphics[width=3.5in]{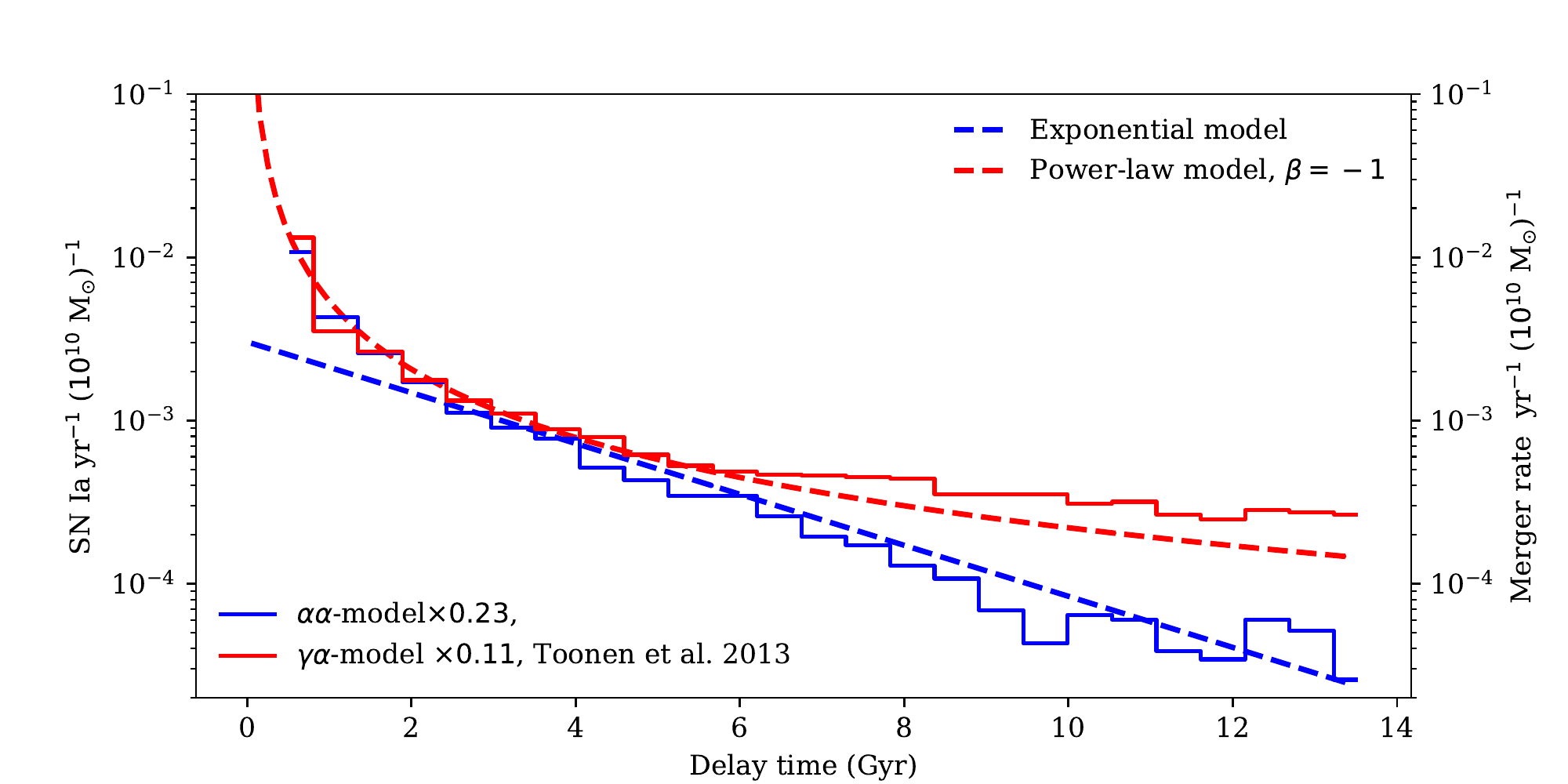}
   \caption{\footnotesize DTDs from this analysis (blue dashed) and $\beta=1$ power-law (red dashed). Also show are the SN~Ia rates as a function of delay time implied from different scenarios for WD/WD binary mergers \citep[see][]{Toonen:2013ng}, scaled to match the model lines. }
   \label{fig:toonen}
\end{figure}

Delay time distribution reconstructions may finally be able to determine whether conservation of orbital energy is the strictly dominant driver (the $\alpha\alpha$-model), as is the case for the common envelope path where both progenitor stars are stripped of their hydrogen envelopes, or if conservation of angular momentum plays a role for most WD/WD mergers initially (the $\gamma\alpha$-model), as it would be in a period of stable mass transfer or in a `formation reversal' evolutionary track \citep{Toonen:2013le}. Figure~\ref{fig:toonen} shows the SN~Ia rate implied from a binary merger rate as a function of delay time for the $\alpha\alpha$-model and $\gamma\alpha$-model of \cite{Toonen:2013ng}, assuming an initial metallicity of $z=0.02$. Also shown are the delay time distributions from the exponential model from this analysis, and the $\beta=-1$ power-law distributions. 

There is striking agreement between the power-law DTD and the $\gamma\alpha$-model (with an appropriate scaling), yet the exponential DTD is more similar to the $\alpha\alpha$-model at large ($>2$ Gyr) delay times. While the implications of these similarities are unclear at this time, it is clear that further refinement of the rate analysis (through improved rate measures) would be fruitful. Similarly, further refinement of the roles of orbital energy and angular momentum conservation in the modeling may be warranted. 

\section{Summary}
We have presented an analysis of type Ia supernova delay time distributions using these two independent methods, from a comparison of volumetric  rates to cosmic star formation histories, and through a maximum likelihood method of host star formation histories and their resultant SN~Ia yields. From this analysis we can conclude the following:
\begin{enumerate}
	\item Volumetric rate measures at or near the same redshift are reasonably consistent with one another, and suggest a collective systematic error of the order of 4 to 17\%.
	\item  Using analytical arguments to fix a value for $k$ (the fraction of stars with initial mass suitable to be SN Ia progenitors), we find the efficiency, $\varepsilon$, of turning those stars into SNe Ia is fairly well constrained at $6.0\pm0.3\%$. This is nearly half the expected value from WD binaries~\citep{Maoz:2017zl}. The combination $k\times\varepsilon$ yields $N/M_\star=1.3^{+0.4}_{-0.3}$ events per 1000 M$_{\odot}$.
	\item The family of delay time distributions solutions we derive from volumetric SN~Ia rates indicate an exponential-like distribution that is somewhat similar to the $\beta\approx-1$ power-law distributions expected from DD progenitor scenarios, and inconsistent with many SD-model expectations from binary population synthesis. 
	\item DTD solutions from host SFHs following the \cite{Maoz:2011} method are identical to those from volumetric rates. 
	\item Exponential solutions are as consistent with empirically recovered delay times as power-law solutions, without having to invoke more than one power-law slope for clustered and field environments.
\end{enumerate}

\acknowledgments We thank our anonymous referee for the highly useful comments and suggestions. Support for HST-GO-14208.006-A was provide by NASA through a grant from the Space Telescope Science Institute, which is operated by the Association of Universities for research in Astronomy, Incorporated, under NASA contract NAS5-26555. O.G. is supported by an NSF Astronomy and Astrophysics Postdoctoral Fellowship under award AST-1602595.

\software{
Astropy \citep{Astropy:2013}, Emcee \citep{Foreman-Mackey:2013pd}, Matplotlib \citep{Hunter:2007pv}, ChainConsumer \citep{Hinton:2016qy}}
\bibliography{strolger}{}

\appendix

\section {Type Ia Supernova Rate Measures}\label{sec:ratemeas}
Table~\ref{tab:sn1a_rates} lists the volumetric SN Ia rate measurements, from various authors, used in the analysis presented in this manuscript.

\startlongtable
\begin{deluxetable}{lcccl}
\label{tab:sn1a_rates}
\tablecaption{Volumetric SN Ia Rates Used in this Work}
\tablehead{
\colhead{Redshift} & \colhead{$R_{\rm Ia}$\tablenotemark{a}} &\colhead{Stat. Uncertainty} & \colhead{Sys. Uncertainty} & \colhead{Source}}

\startdata
0.01&0.28&$^{+0.09}_{-0.09}$&N.A.\tablenotemark{b}&\cite{Cappellaro:1999}\\
0.03&0.28&$^{+0.11}_{-0.11}$&N.A.&\cite{Mannucci:2005}\\
0.0375&0.278&$^{+0.112}_{-0.083}$&$^{+0.015}_{-0.00}$&\cite{Dilday:2010}\\
0.073&0242&$^{0.029}_{-0.029}$&$^{+0.033}_{-0.019}$&\cite{Frohmaier:2019mb}\\
0.1&0.259&$^{+0.052}_{-0.044}$&$^{+0.028}_{-0.001}$&\cite{Dilday:2010}\\
0.10&0.32&$^{+0.15}_{-0.15}$&N.A.&\cite{Madgwick:2003}\\
0.10&0.55&$^{+0.50}_{-0.29}$&$^{+0.20}_{-0.20}$&\cite{Cappellaro:2015oq}\\
0.11&0.37&$^{+0.10}_{-0.10}$&N.A.&\cite{strolger2003}\\
0.13&0.20&$^{+0.07}_{-0.07}$&$^{+0.05}_{-0.05}$&\cite{Blanc:2004}\\
0.15&0.307&$^{+0.038}_{-0.034}$&$^{+0.035}_{-0.005}$&\cite{Dilday:2010}\\
0.15&0.32&$^{+0.23}_{-0.23}$&$^{+0.23}_{-0.06}$&\cite{Rodney:2010b}\\
0.16&0.14&$^{+0.09}_{-0.09}$&$^{+0.06}_{-0.12}$&\cite{Perrett:2012}\\
0.2&0.348&$^{+0.032}_{-0.030}$&$^{+0.082}_{-0.007}$&\cite{Dilday:2010}\\
0.20&0.20&$^{+0.08}_{-0.08}$&N.A.&\cite{Horesh:2008}\\
0.25&0.36&$^{+0.60}_{-0.26}$&$^{+0.12}_{-0.35}$&\cite{Rodney:2014fj}\\
0.25&0.365&$^{+0.031}_{-0.028}$&$^{+0.182}_{-0.012}$&\cite{Dilday:2010}\\
0.25&0.39&$^{+0.13}_{-0.12}$&$^{+0.10}_{-0.10}$&\cite{Cappellaro:2015oq}\\
0.26&0.28&$^{+0.07}_{-0.07}$&$^{+0.06}_{-0.07}$&\cite{Perrett:2012}\\
0.30&0.34&$^{+0.16}_{-0.15}$&N.A.&\cite{Botticella:2008}\\
0.30&0.434&$^{+0.037}_{-0.034}$&$^{+0.396}_{-0.016}$&\cite{Dilday:2010}\\
0.35&0.34&$^{+0.19}_{-0.19}$&$^{+0.19}_{-0.03}$&\cite{Rodney:2010b}\\
0.35&0.36&$^{+0.06}_{-0.06}$&$^{+0.05}_{-0.06}$&\cite{Perrett:2012}\\
0.42&0.46&$^{+0.42}_{-0.32}$&$^{+ 0.10}_{-0.13}$&\cite{Graur:2014}\\
0.44&0.262&$^{+ 0.229}_{-0.133}$&$^{+ 0.059}_{-0.120}$&\cite{Okumura:2014}\\
0.45&0.31&$^{+0.15}_{-0.15}$&$^{+0.15}_{-0.04}$&\cite{Rodney:2010b}\\
0.45&0.36&$^{+0.06}_{-0.06}$&$^{+0.04}_{-0.05}$&\cite{Perrett:2012}\\
0.45&0.52&$^{+0.11}_{-0.13}$&$^{+0.16}_{-0.16}$&\cite{Cappellaro:2015oq}\\
0.46&0.48&$^{+0.17}_{-0.17}$&N.A.&\cite{Tonry:2003}\\
0.47&0.42&$^{+0.06}_{-0.06}$&$^{+0.13}_{-0.09}$&\cite{Neill:2006}\\
0.47&0.80&$^{+0.37}_{-0.27}$&$^{+1.66}_{-0.26}$&\cite{Dahlen:2008}\\
0.55&0.32&$^{+0.14}_{-0.14}$&$^{+0.14}_{-0.07}$&\cite{Rodney:2010b}\\
0.55&0.48&$^{+0.06}_{-0.06}$&$^{+0.04}_{-0.05}$&\cite{Perrett:2012}\\
0.55&0.52&$^{+0.10}_{-0.09}$&N.A.&\cite{Pain:2002}\\
0.65&0.48&$^{+0.05}_{-0.05}$&$^{+0.04}_{-0.06}$&\cite{Perrett:2012}\\
0.65&0.49&$^{+0.17}_{-0.17}$&$^{+0.17}_{-0.08}$&\cite{Rodney:2010b}\\
0.65&0.69&$^{+0.19}_{-0.18}$&$^{+0.27}_{-0.27}$&\cite{Cappellaro:2015oq}\\
0.74&0.79&$^{+0.33}_{-0.41}$&N.A.&\cite{Graur:2011}\\
0.75&0.51&$^{+0.27}_{-0.19}$&$^{+0.23}_{-0.19}$&\cite{Rodney:2014fj}\\
0.75&0.58&$^{+0.06}_{-0.06}$&$^{+0.05}_{-0.07}$&\cite{Perrett:2012}\\
0.75&0.68&$^{+0.21}_{-0.21}$&$^{+0.21}_{-0.14}$&\cite{Rodney:2010b}\\
0.80&0.839&$^{+ 0.230}_{-0.185}$&$^{+ 0.060}_{-0.120}$&\cite{Okumura:2014}\\
0.83&1.30&$^{+0.33}_{-0.27}$&$^{+0.73}_{-0.51}$&\cite{Dahlen:2008}\\
0.85&0.57&$^{+0.05}_{-0.05}$&$^{+0.06}_{-0.07}$&\cite{Perrett:2012}\\
0.85&0.78&$^{+0.22}_{-0.22}$&$^{+0.22}_{-0.16}$&\cite{Rodney:2010b}\\
0.94&0.45&$^{+0.22}_{-0.19}$&$^{+ 0.13}_{-0.06}$&\cite{Graur:2014}\\
0.95&0.76&$^{+0.25}_{-0.25}$&$^{+0.25}_{-0.26}$&\cite{Rodney:2010b}\\
0.95&0.77&$^{+0.08}_{-0.08}$&$^{+0.10}_{-0.12}$&\cite{Perrett:2012}\\
1.05&0.79&$^{+0.28}_{-0.28}$&$^{+0.28}_{-0.41}$&\cite{Rodney:2010b}\\
1.1&0.74&$^{+0.12}_{-0.12}$&$^{+0.10}_{-0.13}$&\cite{Perrett:2012}\\
1.14&0.705&$^{+ 0.239}_{-0.183}$&$^{+ 0.102}_{-0.103}$&\cite{Okumura:2014}\\
1.21&1.32&$^{+0.36}_{-0.29}$&$^{+0.38}_{-0.32}$&\cite{Dahlen:2008}\\
1.23&0.84&$^{+0.25}_{-0.28}$&N.A.&\cite{Graur:2011}\\
1.25&0.64&$^{+0.31}_{-0.22}$&$^{+0.34}_{-0.23}$&\cite{Rodney:2014fj}\\
1.59&0.45&$^{+0.34}_{-0.22}$&$^{+ 0.05}_{-0.09}$&\cite{Graur:2014}\\
1.61&0.42&$^{+0.39}_{-0.23}$&$^{+0.19}_{-0.14}$&\cite{Dahlen:2008}\\
1.69&1.02&$^{+0.54}_{-0.37}$&N.A.&\cite{Graur:2011}\\
1.75&0.72&$^{+0.45}_{-0.30}$&$^{+0.50}_{-0.28}$&\cite{Rodney:2014fj}\\
2.25&0.49&$^{+0.95}_{-0.38}$&$^{+0.45}_{-0.24}$&\cite{Rodney:2014fj}\\
\enddata
\tablenotetext{a}{In units $10^{-4}$ yr$^{-1}$ Mpc$^{-3}$ $h_{70}^3$.}
\tablenotetext{b}{N.A.$=$Not available or cited.}
\end{deluxetable}

\clearpage
\section{MCMC Likelihood Distributions}\label{sec:mcmc_results}
\begin{figure}[t] 
   \centering
   \includegraphics[width=3.1in]{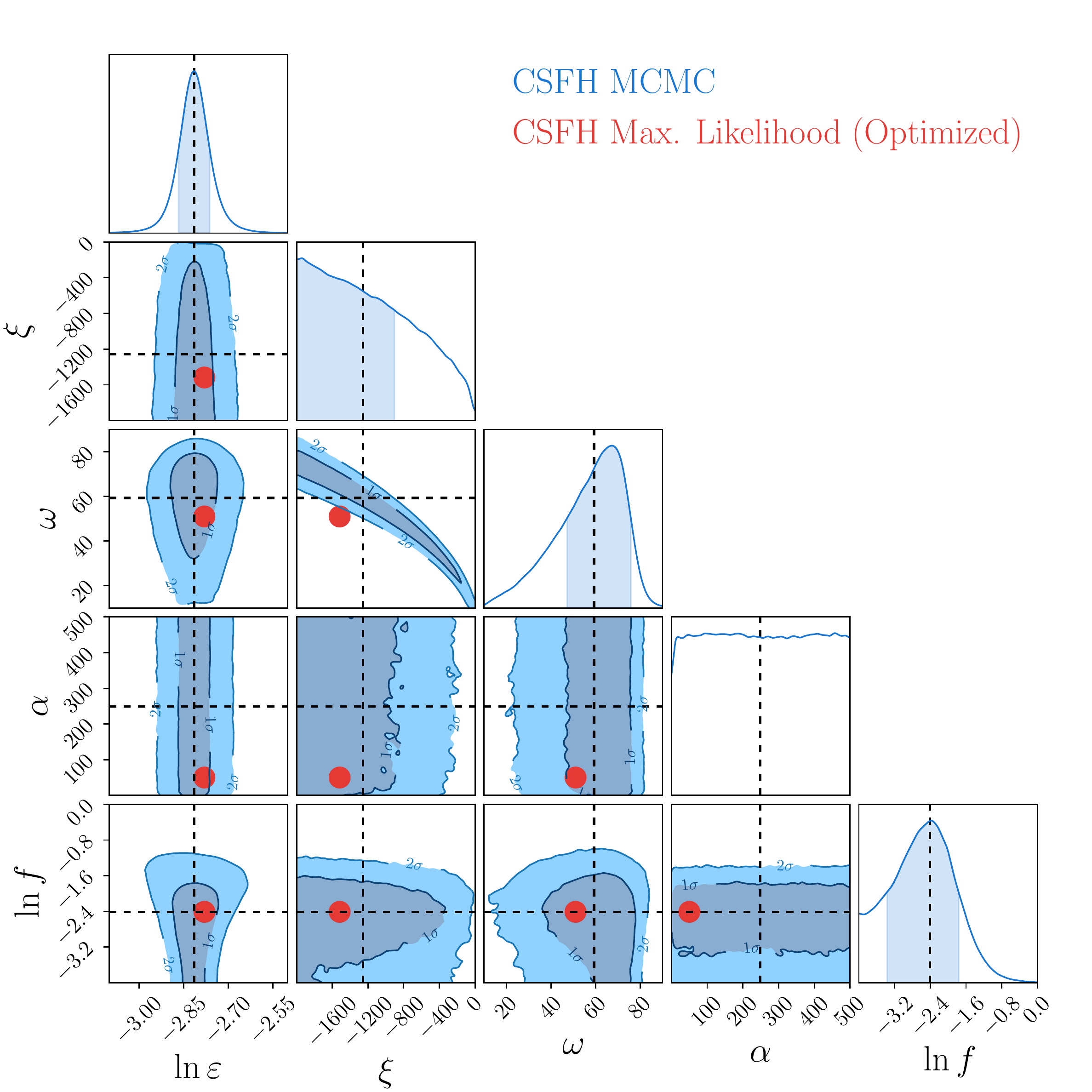} 
   \caption{\footnotesize MCMC results on the parameters of a unimodal delay-time distribution model, fit to volumetric rate data and CSFH. Dashed lines indicate the median values, and the $1-\sigma$ and $2-\sigma$ regions about those best fits are shown in dark and light blue, respectively. The red point marks the maximum likelihood values from the optimized fitting.}
   \label{fig:mcmc_sfd}
\end{figure}

\begin{figure}[t] 
   \centering
   \includegraphics[width=3.1in]{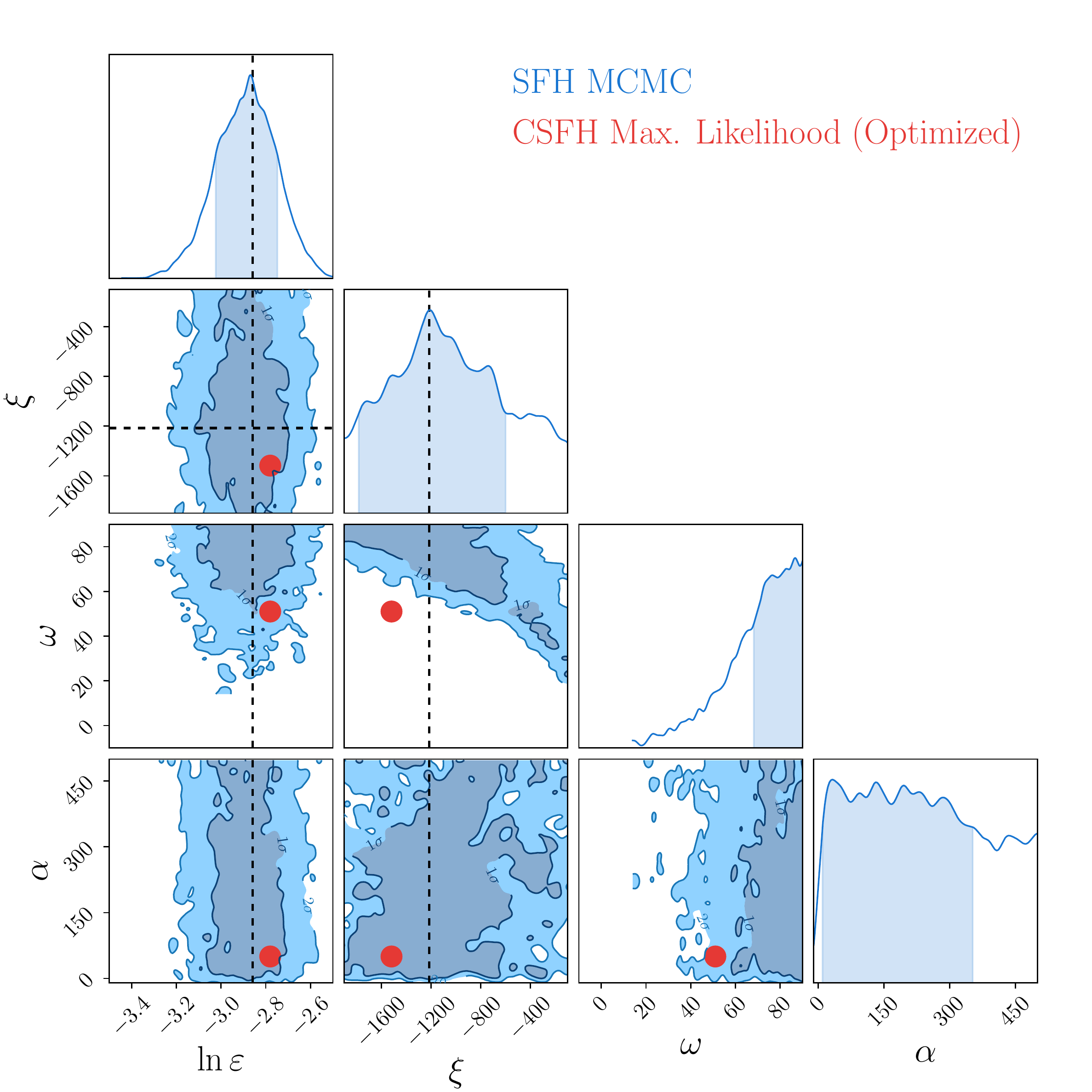} 
   \caption{\footnotesize MCMC results on unimodal delay-time distribution model, fit to SFHs for 68,931 galaxies in the GOODS fields, 67 of which are SN~Ia hosts. Dashed lines indicate the maximum likelihood values. The $1-\sigma$ and $2-\sigma$ regions about those best fits are shown in dark and light blue, respectively. The red point marks the maximum likelihood values from the optimized fitting. }
   \label{fig:mcmc_sfh}
\end{figure}

Shown in Figures~\ref{fig:mcmc_sfd} and \ref{fig:mcmc_sfh} are the MCMC likelihood distributions for the volumetric and individual rates, discussed in Sections~\ref{sec:mcmc_sfd} and \ref{sec:sfh}, respectively. Parameter correlations are shown in Tables~\ref{tab:parameter_correlations1} and \ref{tab:parameter_correlations2}.

\begin{table}[h]
    \centering
    \caption{MCMC CSFH Parameter Correlations}
    \label{tab:parameter_correlations1}
    \begin{tabular}{c|ccccc}
         & $\ln\varepsilon$ & $\xi$ & $\omega$ & $\alpha$ & $\ln f$\\ 
        \hline
        $\ln\varepsilon$ &  1.00 &  0.00 &  0.02 &  0.00 & -0.02 \\ 
                   $\xi$ &  0.00 &  1.00 & -0.95 &  0.01 &  0.02 \\ 
                $\omega$ &  0.02 & -0.95 &  1.00 & -0.01 & -0.05 \\ 
                $\alpha$ &  0.00 &  0.01 & -0.01 &  1.00 &  0.00 \\ 
                 $\ln f$ & -0.02 &  0.02 & -0.05 &  0.00 &  1.00 \\ 
        \hline
    \end{tabular}
\end{table}

\begin{table}[h]
    \centering
    \caption{MCMC SFH Parameter Correlations}
    \label{tab:parameter_correlations2}
    \begin{tabular}{c|cccc}
         & $\ln\varepsilon$ & $\xi$ & $\omega$ & $\alpha$\\ 
        \hline
        $\ln\varepsilon$ &  1.00 & -0.01 &  0.04 & -0.02 \\ 
                   $\xi$ & -0.01 &  1.00 & -0.66 &  0.08 \\ 
                $\omega$ &  0.04 & -0.66 &  1.00 & -0.06 \\ 
                $\alpha$ & -0.02 &  0.08 & -0.06 &  1.00 \\ 
        \hline
    \end{tabular}
\end{table}


\section{The Cosmic Mass-weighted Supernova Rate History}\label{sec:snum}
It can be useful to see what volumetric supernova rates imply for the evolution in mass-weighted supernova rates over cosmic history. Mass-weighted SN rates are often expressed in units of SNuM, or $h^2$ events per century per $10^{10}$ $M_{\odot}$, and are generally convenient for estimating expected yields from individual galaxies. Using observed volumetric rates, they can be found by
\begin{equation}
	{\rm SNuM}(z) = \frac{R(z)}{{\rho}_{\star}(z)}.
\end{equation}
\noindent The evolution of the stellar mass density, ${\rho}_{\star}(z)$, is found by integrating the cosmic star-formation history over time, expressed by
\begin{equation}
	\rho_{\star}(z)=\rho_A\,(1-R) \int\limits_z^{\infty}\frac{\dot{\rho}_{\star}(z')}{H(z')(1+z')}\,dz',
	\label{eqn:smd}
\end{equation}
\noindent where 
\begin{equation}
	H(z) = H_0\sqrt{\Omega_M(1+z)^3+\Omega_{\Lambda}},
\end{equation}
\noindent and $\rho_A=10^{12}\, M_{\odot}$ Mpc$^{-3}$. As is shown in \cite{Madau:2014fk}, the stellar mass density function of Equation~\ref{eqn:smd} matches well to measures from various surveys when the mass fraction of each generation of stars that is put back in to the ISM is $R=0.27$.  

\begin{figure*}[h] 
   \centering
   \includegraphics[width=6.5in]{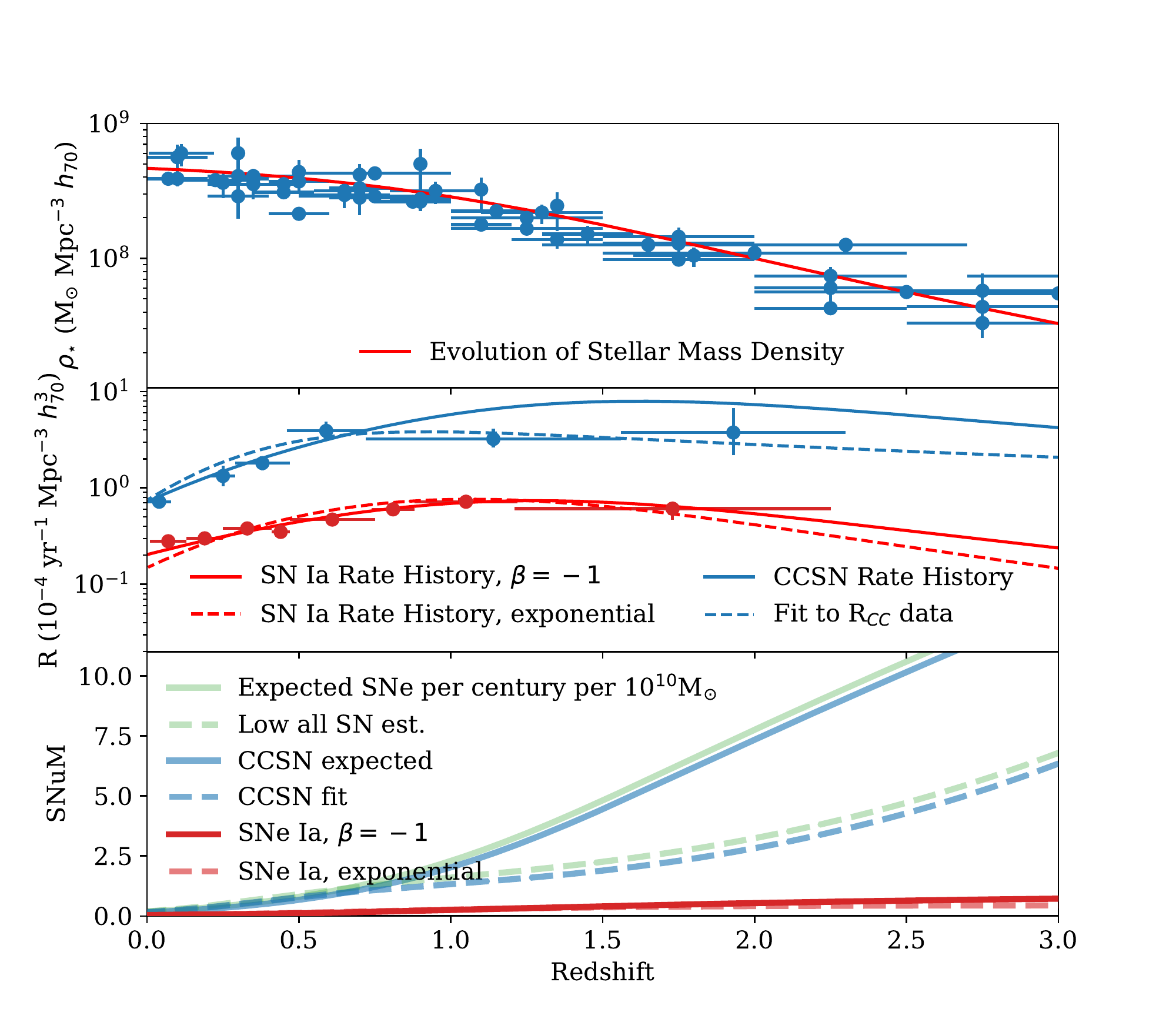}
   \caption{\footnotesize The evolution in mass-weighted SN rates is found by dividing volumetric SN rates by the cosmic evolution of stellar mass density. \textit{Top panel:} the evolution of stellar mass density function in comparison to measures from various authors from the \cite{Madau:2014fk} review. \textit{Middle panel:} volumetric SN rate functions from \cite{Strolger:2015aa} and this manuscript, compared to binned SN rate measures.  \textit{Bottom panel:} resultant SNuM$(z)$ functions.}
   \label{fig:SNuM}
\end{figure*}

Figure~\ref{fig:SNuM} shows the resultant SNuM$(z)$ for function by dividing the stellar mass density function into $R_{\rm Ia}(z)$, for both the $\beta=-1$ power-law and exponential models presented in this paper. Shown also are the results of dividing into the volumetric core-collapse SN rate functions, $R_{\rm cc}(z)$, from \cite{Strolger:2015aa}, assuming either a fit to $R_{\rm CC}$ data or a model which follows a scaled version of the cosmic star formation history. 

\end{document}